\input harvmac.tex
\input amssym.tex

\def\newdate{July 2005}

\def\a{\alpha}
\def\b{\beta}
\def\g{\gamma}
 
\def\d{\delta}
\def\e{\epsilon}
\def\t{\theta}

\def\p{\partial}
\def\half{{1\over 2}}

\def\p{\partial}

\font\cmss=cmss10
\font\cmsss=cmss10 at 7pt
\def\IL{\relax{\rm I\kern-.18em L}}
\def\IH{\relax{\rm I\kern-.18em H}}
\def\IR{\relax{\rm I\kern-.18em R}}
\def\inbar{\vrule height1.5ex width.4pt depth0pt}
\def\IC{\relax\hbox{$\inbar\kern-.3em{\rm C}$}}
\def\rlx{\relax\leavevmode}

\def\ZZ{\rlx\leavevmode\ifmmode\mathchoice{\hbox{\cmss Z\kern-.4em Z}}
 {\hbox{\cmss Z\kern-.4em Z}}{\lower.9pt\hbox{\cmsss Z\kern-.36em Z}}
 {\lower1.2pt\hbox{\cmsss Z\kern-.36em Z}}\else{\cmss Z\kern-.4em
 Z}\fi} 
\def\IZ{\relax\ifmmode\mathchoice
{\hbox{\cmss Z\kern-.4em Z}}{\hbox{\cmss Z\kern-.4em Z}}
{\lower.9pt\hbox{\cmsss Z\kern-.4em Z}}
{\lower1.2pt\hbox{\cmsss Z\kern-.4em Z}}\else{\cmss Z\kern-.4em
Z}\fi}

\def\CN {{\cal N}}


\def\CN {{\cal N}}

\font\manual=manfnt 
\def\dbend{\lower3.5pt\hbox{\manual\char127}}

\def\IZ{\relax\ifmmode\mathchoice
{\hbox{\cmss Z\kern-.4em Z}}{\hbox{\cmss Z\kern-.4em Z}}
{\lower.9pt\hbox{\cmsss Z\kern-.4em Z}} {\lower1.2pt\hbox{\cmsss
Z\kern-.4em Z}}\else{\cmss Z\kern-.4em Z}\fi}
\def\half {{1\over 2}}


\def\p{\partial}

\def\bar{\overline}

\def\CN{{\cal N}}

\def\rt2{\sqrt{2}}
\def\irt2{{1\over\sqrt{2}}}

\def\half{ {1\over 2}}
\def\vp{\varphi}


\lref\Verl{
  H.~Verlinde,
  {\it ``Superstrings on AdS(2) and superconformal matrix quantum mechanics,''}
  arXiv:hep-th/0403024.
}

\lref\BerkovitsZQ{
  N.~Berkovits, M.~Bershadsky, T.~Hauer, S.~Zhukov and B.~Zwiebach,
  {\it ``Superstring theory on AdS(2) x S(2) as a coset supermanifold,''}
  Nucl.\ Phys.\ B {\bf 567}, 61 (2000)
  [arXiv:hep-th/9907200].
}

\lref\BerkovitsVN{
  N.~Berkovits and J.~Maldacena,
  {\it ``N = 2 superconformal description of superstring in Ramond-Ramond plane wave backgrounds,''}
  JHEP {\bf 0210}, 059 (2002)
  [arXiv:hep-th/0208092].
}
\lref\BerkovitsDU{
  N.~Berkovits,
  {\it ``Quantization of the type II superstring in a curved six-dimensional
  background,''}
  Nucl.\ Phys.\ B {\bf 565}, 333 (2000)
  [arXiv:hep-th/9908041].
}

\lref\BerkovitsWR{
  N.~Berkovits,
  {\it ``Covariant quantization of the Green-Schwarz superstring in a Calabi-Yau
  background,''}
  Nucl.\ Phys.\ B {\bf 431}, 258 (1994)
  [arXiv:hep-th/9404162].
}

\lref\BerkovitsVY{
  N.~Berkovits and C.~Vafa,
 {\it  ``N=4 topological strings,''}
  Nucl.\ Phys.\ B {\bf 433}, 123 (1995)
  [arXiv:hep-th/9407190].
}

\lref\BerkovitsIM{
  N.~Berkovits, C.~Vafa and E.~Witten,
 {\it  ``Conformal field theory of AdS background with Ramond-Ramond flux,''}
  JHEP {\bf 9903}, 018 (1999)
  [arXiv:hep-th/9902098].
}

\lref\BerkovitsDU{
  N.~Berkovits,
 {\it  ``Quantization of the type II superstring in a curved six-dimensional
  background,''}
  Nucl.\ Phys.\ B {\bf 565}, 333 (2000)
  [arXiv:hep-th/9908041].
}

\lref\BerkovitsBF{
  N.~Berkovits,
  {\it ``A new description of the superstring,''}
  arXiv:hep-th/9604123.
}

\lref\BerkovitsCB{
  N.~Berkovits and W.~Siegel,
  {\it ``Superspace Effective Actions for 4D Compactifications of Heterotic and Type II Superstrings,''}
  Nucl.\ Phys.\ B {\bf 462}, 213 (1996)
  [arXiv:hep-th/9510106].
}

\lref\BerkovitsVN{
  N.~Berkovits and J.~Maldacena,
  {\it ``N = 2 superconformal description of superstring in Ramond-Ramond plane  wave backgrounds,''}
  JHEP {\bf 0210}, 059 (2002)
  [arXiv:hep-th/0208092].
}

\lref\kutsei{D.~Kutasov and N.~Seiberg, 
{\it ``Non-critical superstrings''}, Phys.\ Lett.\ B {\bf 251}, 67 (1990).}

\lref\kutasov{D.~Kutasov, 
{\it ``Some properties of (non)critical strings''}, [arXiv:hep-th/9110041].}

\lref\PolyakovRE{
  A.~M.~Polyakov,
 {\it  ``Quantum Geometry Of Fermionic Strings,''}
  Phys.\ Lett.\ B {\bf 103}, 211 (1981).
}

\lref\PolyakovRD{
  A.~M.~Polyakov,
 {\it  ``Quantum Geometry Of Bosonic Strings,''}
  Phys.\ Lett.\ B {\bf 103}, 207 (1981).
}

\lref\PolyakovFK{
  A.~M.~Polyakov,
 {\it  ``String theory as a universal language,''}
  Phys.\ Atom.\ Nucl.\  {\bf 64}, 540 (2001)
  [Yad.\ Fiz.\  {\bf 64}, 594 (2001\ IMPAE,A16,4511-4526.2001)]
  [arXiv:hep-th/0006132].
}

\lref\KupersteinYK{
  S.~Kuperstein and J.~Sonnenschein,
  {\it ``Non-critical supergravity ($d > 1$) and holography,''}
  JHEP {\bf 0407}, 049 (2004)
  [arXiv:hep-th/0403254],
``Non-critical, near extremal AdS(6) background as a holographic laboratory
  of four dimensional YM theory,''
  JHEP {\bf 0411}, 026 (2004)
  [arXiv:hep-th/0411009].
}

\lref\KlebanovYA{
  I.~R.~Klebanov and J.~M.~Maldacena,
  {\it ``Superconformal gauge theories and non-critical superstrings,''}
  Int.\ J.\ Mod.\ Phys.\ A {\bf 19}, 5003 (2004)
  [arXiv:hep-th/0409133].
}

\lref\BerkovitsFE{
  N.~Berkovits,
  {\it ``Super-Poincare covariant quantization of the superstring,''}
  JHEP {\bf 0004}, 018 (2000)
  [arXiv:hep-th/0001035].
}

\lref\ItohQB{
  K.~Itoh and N.~Ohta,
  {\it ``BRST cohomology and physical states in 2-D supergravity coupled to c $\leq$ 1
  matter,''}
  Nucl.\ Phys.\ B {\bf 377}, 113 (1992)
  [arXiv:hep-th/9110013].
}

\lref\BouwknegtAM{
  P.~Bouwknegt, J.~G.~McCarthy and K.~Pilch,
  {\it ``Ground ring for the 2-D RNS string,''}
  Nucl.\ Phys.\ B {\bf 377}, 541 (1992)
  [arXiv:hep-th/9112036].
}

\lref\BouwknegtVA{
  P.~Bouwknegt, J.~G.~McCarthy and K.~Pilch,
  {\it ``BRST analysis of physical states for 2-D (super)gravity coupled to
  (super)conformal matter,''}
  arXiv:hep-th/9110031.
}

\lref\MurthyES{
  S.~Murthy,
  {\it ``Notes on non-critical superstrings in various dimensions,''}
  JHEP {\bf 0311}, 056 (2003)
  [arXiv:hep-th/0305197].
}

\lref\PolyakovBR{
  A.~M.~Polyakov,
 {\it ``Conformal fixed points of unidentified gauge theories,''}
  Mod.\ Phys.\ Lett.\ A {\bf 19}, 1649 (2004)
  [arXiv:hep-th/0405106].
}

\lref\ItaNE{
  H.~Ita, H.~Nieder and Y.~Oz,
  {\it ``On type II strings in two dimensions,''}
  arXiv:hep-th/0502187.
}

\lref\SeibergBX{
  N.~Seiberg,
  {\it ``Observations on the moduli space of two dimensional string theory,''}
  arXiv:hep-th/0502156.
}

\lref\tama{
  P.~A.~Grassi and L.~Tamassia,
  {\it ``Vertex operators for closed superstrings,''}
  JHEP {\bf 0407}, 071 (2004)
  [arXiv:hep-th/0405072].
}
\lref\PolyakovAM{
  A.~M.~Polyakov and A.~B.~Zamolodchikov,
  {\it ``Fractal Structure Of Two-Dimensional Supergravity,''}
  Mod.\ Phys.\ Lett.\ A {\bf 3}, 1213 (1988).
}

\lref\GiveonZM{
  A.~Giveon, D.~Kutasov and O.~Pelc,
  {\it ``Holography for non-critical superstrings,''}
  JHEP {\bf 9910}, 035 (1999)
  [arXiv:hep-th/9907178].
}

\lref\BerkovitsTG{
  N.~Berkovits, S.~Gukov and B.~C.~Vallilo,
  {\it ``Superstrings in 2D backgrounds with R-R flux and new extremal 
  black  holes,''}
  Nucl.\ Phys.\ B {\bf 614}, 195 (2001)
  [arXiv:hep-th/0107140].
}

\lref\BigazziMD{
  F.~Bigazzi, R.~Casero, A.~L.~Cotrone, E.~Kiritsis and A.~Paredes,
  {\it ``Non-critical holography and four-dimensional CFT's with fundamentals,''}
  arXiv:hep-th/0505140.
}

\lref\AlishahihaYV{
  M.~Alishahiha, A.~Ghodsi and A.~E.~Mosaffa,
  {\it ``On  isolated conformal fixed points and noncritical string theory,''}
  JHEP {\bf 0501}, 017 (2005)
  [arXiv:hep-th/0411087].
}


\Title{\vbox{
\hbox{CERN-PH-TH/2005-061}
\hbox{TAUP-2801-05}
}}
{\vbox{\centerline{
Non-Critical Covariant Superstrings}
}}
\smallskip
\medskip
\centerline{
{Pietro~Antonio~Grassi}$^{~a,b,c,}$\foot{pgrassi@cern.ch} 
and {Yaron~Oz}$^{~d,}$\foot{yaronoz@post.tau.ac.il}}
\bigskip 
\centerline{\it $^{a}$ CERN, Theory Division, CH-1121 Genev\'e 23, Switzerland,}  
\centerline{\it $^{b}$ DISTA, Universit\`a del Piemonte Orientale,}
\centerline{\it ~~~~ Via Bellini, 25/g 15100 Alessandria, Italy,}
\centerline{\it $^{c}$ Centro E. Fermi, Compendio Viminale, I-00184, Roma, Italy,} 
\centerline{\it $^{d}$ Raymond and Beverly Sackler Faculty of Exact Science, }
\centerline{\it School of Physics and Astronomy, Tel-Aviv University, Ramat-Aviv 69978, Israel} 

\bigskip
\bigskip

We construct a covariant description of 
non-critical superstrings in even dimensions. 
We construct explicitly supersymmetric
hybrid type variables in a linear dilaton background, and study an underlying
$N=2$ twisted superconformal algebra structure. We find similarities between
non-critical superstrings in $2n+2$ dimensions and
critical superstrings compactified on $CY_{4-n}$ manifolds.
We study the 
spectrum 
of the non-critical strings, and in particular the Ramond-Ramond
massless fields.
We use the  supersymmetric
variables to construct the  non-critical superstrings $\sigma$-model
action
in curved target space backgrounds
with coupling to the Ramond-Ramond fields.
We consider as an example  non-critical
type IIA strings  on $AdS_{2}$ background with Ramond-Ramond 2-form flux.


\noindent
 \Date{\newdate}

\listtoc
\writetoc
\vfill
\eject


\newsec{Introduction}

The critical dimension for  superstrings in flat space is $d=10$.
In dimensions $d<10$, the Liouville mode is dynamical and should
be quantized as well \refs{\PolyakovRE,\PolyakovRD}. 
Such strings are called non-critical.
The total conformal anomaly vanishes  for the non-critical
strings due to the Liouville background charge.
Our aim is to construct a manifestly supersymmetric and 
covariant worldsheet description of non-critical superstrings, 
in particular for 
curved  target space geometries and with coupling to Ramond-Ramond fields.

There are various  motivations to study
non-critical  strings.
First, non-critical  strings can provide an alternative
to string compactifications.
Second,  non-critical  strings can  provide a dual description
of gauge theories.
Examples of backgrounds one wishes to study
are  \PolyakovFK
\eqn\metric{
ds^2 = d \varphi^2 + a^2(\varphi)d \vec{x}^2 \ ,
}
where $\vec{x}=(x_1,...,x_{d-1})$, and with other background fields 
turned on. String theory on such warped backgrounds
is expected
to provide a dual description of gauge theories.
Depending on the form of the warp factor  $a^2(\varphi)$, the gauge theory
can be confining, or at
a conformal fixed point.

One complication in the study of  non-critical superstrings
is that unlike the critical case, there is
no consistent approximation where  supergravity provides
a valid effective description.
The reason being that the $d$-dimensional  
supergravity low-energy effective action
contains a cosmological constant type term of the
form
\eqn\cosmo{
S \sim \int d^d x \sqrt{G}e^{-2 \Phi}\left({d-10 \over l_s^2}\right)\,,
}
which vanishes only for $d=10$.
This implies that the
low energy approximation $E\ll l_s^{-1}$ is not valid when $d \neq 10$,
and the higher order curvature terms of the form $\left(l_s^2{\cal R}\right)^n$ 
cannot be discarded.
A manifestation of this is that solutions of the  $d$-dimensional
supergravity equations have typically curvatures of the 
order of the string scale  $l_s^2{\cal R}\sim O(1)$
when  $d \neq 10$ \refs{
\KupersteinYK,\PolyakovBR,\KlebanovYA,\AlishahihaYV,\BigazziMD}.

The second complication is that interesting target space
curved geometries include  Ramond-Ramond (RR) field fluxes, and we face
the need to quantize the strings in such backgrounds.
The conventional (Ramond-Neveu-Schwarz) 
RNS  formalism is inadequate, since it does not
treat the RR fields on the same footing as the (Neveu-Schwarz) 
NSNS fields due to 
the non-polynomial couplings of the RR fields to the 
spin fields of the CFT. A 
framework to study curved geometries that include RR fields fluxes,
is a target space covariant formulation
of non-critical superstrings. 
We expect this also to enable us to study D-branes in the 
non-critical superstrings and their dual configurations in the gauge theories.

We will start by considering the  tachyon free non-critical superstrings of
\kutsei.
These  are  $(2n+2)$-dimensional fermionic strings with  $n=0,1,2,3$ 
($n=4$ is the
ten-dimensional critical superstring).
The target space geometry is flat with a linear dilaton field. 
In the superconformal gauge they are described
by $2n +1$ matter superfields
$X^i,i=1, \dots, 2n+1 \equiv D$, and by a Liouville superfield $\Phi_l$.
In components we have  $(x^{i}, \psi^{i})$, $(\varphi, \psi_l)$.
We will build supersymmetric variables and use them  
to construct a covariant description of these non-critical superstrings.
We will show how the RNS GSO projection is implemented automatically
in 
the  covariant formalism, and  
use this structure to couple the $\sigma$-model to curved backgrounds
with RR fields. 
Unlike the Green-Schwarz  (GS) $\sigma$-model construction, we will work
in the framework where $\kappa$-symmetry is already fixed. It is also 
important to stress another fundamental difference compared
to the GS formulation which 
is the field content of the $\sigma$-model. Besides 
the usual bosonic coordinates 
and the fermionic superspace coordinates $\t$'s, in the present 
formulation there are also the conjugate momenta to $\t$'s, denoted here 
by $p$'s which provide the linear couplings to the Ramond fields. 

The space-time supersymmetry of the 
$(2n+2)$-dimensional strings has effectively a supersymmetry
structure of  $2n$-dimensional space-time.
In the RNS formalism  of a linear dilaton background
only half of the supercharges  which can be 
constructed are mutually local with respect to each other
\kutasov, and only half of the  mutually local supercharges
are actual supersymmetries.
However, 
to construct a manifest superspace approach to these string 
theories we will use
a bigger superspace with a double number 
of fermionic coordinates with respect to the number of supersymmetries 
manifest. 
This superspace structure suggests that the
non-critical superstrings may have solutions with double
the number of supersymmetries of the linear dilaton background.
We then follow a similar prescription to that of the 
hybrid formalism to build  the worldsheet 
$N=2$ superconformal generators in a 
manifestly space-time  supersymmetric way.
The basic algebraic structure that we will use
to compare the RNS variables to the hybrid type variables
is
a $\hat{c}=2$ twisted $N=2$ superconformal algebra, 
where the dimension one current is the BRST current of the 
non-critical superstring. 

We will find a similar algebraic structure for
the $(2n+2)$-dimensional fermionic non-critical 
superstrings with  $n=0,1,2,3$
and the compactification independent part of 
critical fermionic strings compactified on a $CY_{4-n}$
manifold.
More precisely, we will identify an underlying 
$\hat{c}=n-2$ twisted $N=2$ superconformal algebra.
Note, however, that the two systems are different. In particular, they differ
by the amount of space-time supersymmetry, and the RR fields.

The paper is organized as follows.
In the next 
section we review the structure of the non-critical strings in the RNS formalism.
We analyze the BRST symmetry, $N=2$ superconformal symmetry,  the space-time
supersymmetry
 and the spectrum.
We construct a  $\hat{c}=2$ twisted $N=2$ superconformal algebra that will be used as
a basic structure to compare with the hybrid formalism.
In section 3 we construct the covariant non-critical superstrings using 
hybrid variables. 
 We compare the twisted $N=2$ superconformal algebra
with the RNS one.
We consider in detail the two-dimensional superstrings. 
We construct the BRST operators and write the spectrum at ghost number one as well as
the ground ring generators at ghost number zero in the supersymmetric
variables. We show how the RNS GSO projection is implemented in 
the  covariant formalism.
We identify an underlying 
$\hat{c}=-2$ twisted $N=2$ superconformal algebra, that compares with
that of critical fermionic strings compactified on a $CY_4$
manifold.
In section 4 we construct the covariant higher dimensional 
 non-critical superstrings. 
The structure is similar to that of the two-dimensional
superstrings.
Again, we will enlarge the superspace and compare with 
 critical fermionic strings compactified on $CY$
manifolds.
We will comment on some subtleties in the cases $n=2$ and $n=3$ associated
with the structure of the hybrid formalism. These
cases work out much like   critical fermionic strings compactified on a 
$CY_2$ and $CY_1$, respectively.
In section 5  we analyze  non-critical strings 
in curved target space backgrounds
with coupling to Ramond-Ramond fields.
We construct the worldsheet $\sigma$-models and  
consider the example of non-critical type IIA 
strings on $AdS_{2}$ background with Ramond-Ramond 2-form fields flux.
Section 6 is devoted to a discussion.

\newsec{Non-critical Superstrings}

In this section we will consider  fermionic strings
propagating on a linear dilaton background
of the form (in the string frame)
\eqn\background{
ds^2 = \eta_{ij}dx^i dx^j + dx^2 +d\varphi^2,~~~~~\Phi = -{Q \over \sqrt{2}}\varphi\,,
}
where $i,j=1,...,2n, n=0,1,2,3$, $x \in S^1$ of radius $R={2/Q}$
($\alpha' = 2$),  $Q=\sqrt{4-n}$, 
and $\Phi$ is the dilaton field. 
Note that $n=4$ is the critical superstrings case.
The analysis will be performed
in the perturbative 
regime where the string coupling is small: $g_s \sim e^{\Phi}
\sim e^{-{Q \over \sqrt{2}}\varphi}\ll 1$.

Since a flat background with constant dilaton field is not a solution
of the non-critical string equations, the  linear dilaton background
will be used by us to make the dictionary between the RNS non-critical
strings and the covariant   non-critical strings.
This dictionary will be used later in order to couple the non-critical
strings to curved backgrounds with RR fields.

\subsec{RNS variables}

We start by discussing the non-critical fermionic 
strings in the RNS 
formalism \kutsei.
The $(2n+2)$-dimensional fermionic strings with  $n=0,1,2,3$, are
described in the superconformal gauge by $2n +1$ matter superfields
$X^i,i=1, \dots, 2n +1 \equiv D$, and by a Liouville superfield $\Phi_l$.
In components we have  $(x^{i}, \psi^{i})$, $(\varphi, \psi_l)$, where
 $\psi^{i}$ and $\psi_l$ are Majorana-Weyl fermions.
As usual, we have two ghost systems
$(\beta, \gamma)$ and $(b,c)$.
The central charges  are given by
 $(2n+1, (2n+1)/2)$, $(1 + 3 Q^{2},1/2)$, $11$ and $-26$.
 The total central charge is 
given by $ 3(2n+1)/2 + 1/2 + (1 + 3 Q^{2}) -15 = 3 ( n + Q^{2} - 4)$. It 
vanishes for 
\eqn\cc{Q(n) = \sqrt{4-n} \ .
}
For $n=4$, we have nine flat  coordinates 
and together with the Liouville field and $x$, this gives
 the flat ten dimensional
critical superstring.
When  $n\neq 4$ we  call the resulting systems non-critical superstrings.

The stress energy tensor of the system reads  
\eqn\conA{
T= T_{m} + T_{ghost}\,,
}
$$
T_{m} = 
\sum^{2n +1}_{i=1} 
\Big( - \half (\p x^{i})^{2} - \half \psi^{i} \p \psi^{i} \Big) + 
\Big( - \half  (\p \vp )^{2} + {Q(n)\over 2} \p^{2} \vp - 
\half \psi_l \p \psi_l \Big)  
$$
$$
T_{ghost} = - 2 b \p c - \p b c - {3\over 2} \b \p \g - {1\over 2} \p \b \g\,, 
$$

The OPE's conventions that we will be using are 
$$
x^{i}(z) x^{j}(w) \sim - \eta^{ij} \log(z-w) \,, ~~~~
\vp(z)  \vp(w) \sim - \log(z-w)\,, ~~~
$$
$$
\psi^{i} \psi^{j} \sim \eta^{ij} {1\over (z-w)}\,, ~~~~~
\psi_{l} \psi_{l} \sim {1\over (z-w)}\,,
$$
$$
c(z) b(w) \sim {1\over (z -w)} \,, ~~~~
\gamma(z) \beta(w) \sim {1\over (z-w)}\,,
$$
$$
T(z) e^{r x}(w) \sim 
\Big( { r^{2}/2 \over (z-w)^{2}} + 
{\partial \over (z-w)} \Big)
e^{r x}\,, ~~~~
T(z) e^{s \vp}(w) \sim 
\Big( {- s(s - Q(n))/2 \over (z-w)^{2} } + 
{\partial \over (z-w)} \Big)
e^{s \vp}\,.
$$

We choose an euclidean metric $\eta^{ij}$
for the bosonic space $x^{i}$. However, 
one of the boson, which we take to be $x^{2n+1} \equiv x$ is
compactified on $S^{1}$ of radius ${2 \over Q}$.
In this case the system has global $N=2$ symmetry on the worldsheet,
which will be discussed in the next section.

We define $\Psi = \psi_l + i \psi$ and 
$\Psi^{I} = \psi^{i} + i \psi^{i+n}$ (with $I = 1, \dots, n$). These are bosonized in the usual way by introducing the bosonic fields $H$ for 
$\p H = {i \over 2}\Psi \Psi^{\dagger}$ and 
$\p H^{I} = {i \over 2}\Psi^{I} \Psi^{\dagger I}$ where  
$\Psi^{\dagger} = \psi - i \psi_{l}$,
$\Psi^{\dagger I} = \psi^{i} - i \psi^{i+n}$
is complex conjugation in field space.
We have 
\eqn\H{ 
H^{i}(z) H^{i}(w) \sim - \log(z-w) \,, ~~~~
H(z) H(w) \sim - \log(z-w) \ .
}
We define the spin fields $\Sigma^{\pm} = 
e^{{\pm } {i \over 2} H}$.  In addition, we define the other spin fields 
$\Sigma^{a} = e^{\pm {i \over 2}H^{1}  \dots \pm {i \over 2}
H^{n}}$, where the index $a$ 
runs over the independent spinor representation of $SO(2n)$.
We have $H^{\dagger} = H$ and  $H^{\dagger I} = H^I$.


\subsec{BRST symmetry and N=2 superconformal symmetries}

The matter and Liouville system has
$N=2$ global superconformal 
symmetry with central charge $\hat c = 1 + n + Q^{2}$ 
generated by
\eqn\scC{
G^{+} = 
{1\over 2} \sum_{I=1}^{n} \Psi_{I}^{\dagger} \p (x_{I} + i x_{I +n}) + 
{1\over 2} \Psi^{\dagger} \p (\vp + i x) - 
{1\over 2} Q(n)  \p \Psi^{\dagger}\,.
}
$$
G^{-} = 
{1\over 2} \sum_{I=1}^{n} \Psi_{I}  \p (x_{I} - i x_{I +n}) + 
{1\over 2} \Psi \p (\vp - i x) -  
{1\over 2} Q(n)  \p \Psi\,.
$$
$$
J = 
{1\over 2} \sum_{I} \Psi^{I} \Psi^{\dagger I} + 
 {1\over 2} \Psi \Psi^{\dagger} + 
i Q(n) \p x\,,
$$
and the energy momentum tensor $T$ given in \conA. 
We have $T^{\dagger} = T,
(G^{\pm})^{\dagger} = G^{\mp}, J^{\dagger} = -J$.  
 
The central charge can be computed by the central term of $J$
\eqn\J{
J(z) J(w) \sim  {\hat c \over (z-w)^2} \ ,
}
and 
is $\hat c = n + 1 + Q^{2}(n) = 5$ for $Q(n)$ 
given in \cc. 

To pick an $N=1$ algebra to gauge inside the $N=2$ algebra, we 
consider the combination $G_m= {1\over \sqrt{2}} (G^{+} + G^{-})$   
as the $N=1$ supersymmetry generator  for the matter system.
Note that  $G^{\dagger}_m = G_m$. 
There are other choices for picking the
supersymmetry generator. 
They are parametrized
by $SU(2)/U(1)$ ways of choosing the $N=1$ algebra inside 
the $N=2$ algebra
of \scC.

We define a twisted N=2 superconformal algebra 
by the generators 
\eqn\scA{
G^{\prime +} = 
\gamma G_{m} + c \Big(T_{m} - {3\over 2} \beta \p \gamma - {1\over 2} \gamma \p\b - b\p c\Big) - 
\g^{2} b + 
\p^{2}c + 
\p(c \xi \eta) \,, ~~~
}
$$
G^{\prime -} = b\,, ~~~~~~~
J^{\prime} = c b + \eta \xi\,, ~~~~~~
T^{\prime} =  T_{m} + T_{ghost}\,.
$$

The dimension one current 
$G^{\prime +}$ is the BRST current of the superstring. 
$\xi$ and $\eta$ are defined via the bosonization of
the superghosts
$\gamma = e^{\phi} \eta, \beta = \p \xi  e^{-\phi}$, as will be discussed
shortly, and $J^{\prime}$ is the ghost current.
Notice also that the last two terms in the BRST current do not affect the BRST charge, 
but only the BRST current in such a way that $G^{\prime +}(z) G^{\prime +}(w) \rightarrow 0$. 
The central charge of the twisted N=2 superconformal algebra
\scA\ can be computed by 
the central term of $J'$, which gives $\hat c =2$. 

Let us bosonize the ghost systems.  We define  
\eqn\bc{
c = e^{\chi}\,,~~~~~ b = e^{-\chi}\,,
}
with
\eqn\conB{
T_{b,c}  = T_{\chi}= \half (\p \chi)^{2} + {3\over 2} \p^{2} \chi\,, ~~~~
c(z) b(w) \sim {1\over z-w} + \p \chi\,,~~~~
}
$$
\chi(z) \chi(w) \sim \log(z-w) \,, ~~~~~
$$
$$
T_{\chi}(z) \p \chi(w) \sim {-3 \over (z-w)^{3}} + \dots\,, ~~~~
 T_{\chi}(z) e^{a \chi}(w) \sim 
 \Big( 
 {a(a - 3)/2 \over (z-w)^{2}} +  {\p \over (z-w)} 
 \Big) e^{a\chi} \,. 
$$
The background charge is $Q_{b,c} = 3$ and the total charge 
is $(1 + 3 \e Q^{2}) = -26$ with $\e = \pm$ for bosons/fermions. 

For the superghosts, 
we have $\gamma = e^{\phi} \eta, \beta = \p \xi  e^{-\phi}$
\eqn\coC{
T_{\b,\g} = T_{\phi} + T_{\eta,\xi} = -\half (\p \phi)^{2} - \p^{2} \phi  - \eta \p\xi\,,~~~~ \g(z) \b(w) \sim {1\over (z-w)} + \p \phi + \dots\,,
}
$$
\phi(z) \phi(w) \sim {- \log(z-w)}\,, ~~~~ \eta(z) \xi(w) \sim {1\over (z-w)}\,, ~~~
$$
$$
T_{\phi} \p\phi \sim {2\over (z-w)^{3}} + \dots\,, ~~~~
T_{\phi}(z) e^{b \phi}(w) \sim 
 \Big( 
 {-b(b + 2)/2 \over (z-w)^{2}} +  {\p \over (z-w)} 
 \Big) e^{b\phi} \,.
$$
The background charge is $Q_{\phi} = - 2$ and the total conformal 
charge is $(1 + 3 Q^{2}) -2 = 13 -2 =11$. 
We further bosonize the fermions into 
\eqn\exi{
\eta = e^{\kappa}\,,~~~~~ \xi=e^{- \kappa}\,,
}
and 
\eqn\coD{
T_{\eta,\xi}  = T_{\kappa} = 
\half (\p \kappa)^{2} - \half \p^{2} \kappa\,, ~~~~
\eta(z) \xi(w) \sim {1\over (z-w)} + \p \kappa + \dots\,.
}
$$
\kappa(z) \kappa(w) \sim \log(z-w)\,,
$$
$$
T_{\kappa} \p\kappa \sim {1\over (z-w)^{3}} + \dots\,, ~~~~
T_{\kappa}(z) e^{c \kappa}(w) \sim 
 \Big( 
 {c(c +1)/2 \over (z-w)^{2}} +  {\p \over (z-w)} 
 \Big) e^{c\kappa} \,.
$$
The background charge is $Q = -1$ and the total charge is 
$(1 - 3Q^{2}) = - 2$. So, finally we use the ghosts 
$\gamma = e^{\phi + \kappa}$ and $\beta =  \p \kappa e^{- \phi - \kappa}$. 

\subsec{Supersymmetry}

In the following we will discuss the supersymmetry structure
in the RNS formalism (see \kutsei\ and also \refs{\kutasov,\MurthyES,\SeibergBX,\ItaNE}),
and 
how it will be realized in terms of the supersymmetric variables. 
For the $(2n+2)$-dimensional strings we can construct in the $-\half$ picture
$2^{n+2}$ candidates for supercurrents
\eqn\currents{
e^{-{\phi \over 2} + 
{i \over 2}\left(\pm H \pm H^1 \pm ...\pm H^n \pm Q x \right)}\,.
}  
However, only $2^n$ of them are mutually local w.r.t each other
and close a supersymmetry algebra.
Combining the left and right sectors, 
one gets an $N=2$ supersymmetry algebra in $2n$-dimensional space.
Type IIA and type IIB strings are distinguished in the way we choose the
supersymmetry currents from the 
left and right sectors.
When the target space allows chiral supersymmetry ($n=1,3$), type IIA
and  type IIB have 
$(1,1)$  and $(2,0)$   target space supersymmetry, respectively.
In order to work in a covariant formalism we will see that it is convenient
to use a bigger superspace with double the amount of supersymmetric
coordinates, namely $2^{n+1}$ supersymmetric coordinates
from the left sector and  $2^{n+1}$ supersymmetric coordinates
from the right sector. 
Such superspace arises naturally when considering 
the critical superstrings compactified on
$CY_{4-n}$ manifolds.
It suggests that the non-critical superstrings may have
solutions with double the supersymmetry
of the linear dilaton background.
Let us see how this works in detail.

We start from the simplest model with $D=1$ 
($n=0$). We have the bosonic fields $(x,\varphi)$.
In this case there is only one nilpotent supercharge.
We can choose the corresponding supercurrent 
$q_+(z)$ in the form 
\eqn\super{
q_+(z)=e^{-{\phi \over 2} -{iH \over 2}-ix} \,.
}
$\phi$, $H$ and $x$ are holomorphic parts of scalar fields.

The supercharge $Q_{+}$ is given by
\eqn\Q{
Q_{+}  = \oint e^{- {1\over 2} \phi - {i\over 2} H - i x}\,,
}
with $Q_{+}^2=0$.

One can write another supercurrent in the form
\eqn\supernext{
q_-(z)= e^{-{\phi \over 2} + {iH \over 2}+ ix}\,.
}
However, it is not 
local w.r.t.  $q_+(z)$. 
We have $(Q_{+})^\dagger = Q_{-}$.

There is also a
supercurrent from  the right sector, which we will denote by
$\bar{q}$. If we choose, the same supercurrents
$(q_+,\bar{q}_+)$ or $(q_-,\bar{q}_-)$
in the left and right sector, we get type IIB with $0+0$-dimensional
$N=2$ supersymmetry (the two supercharges are nilpotent).
If, on the other hand, we choose different supercurrents in the left and right sectors
$(q_+,\bar{q}_-)$ or $(q_-,\bar{q}_+)$
we get type IIA with $0+0$-dimensional
$N=2$ supersymmetry, again with two nilpotent supercharges.
The affine current
\eqn\J{
J_R =  {2 i \over Q(n)} \p x \,,
}
corresponds to the $U(1)_R$ symmetry.
$q_+$ and $q_-$ have R-charges $\pm 1$. 
Note that while the target space is two-dimensional
with coordinates $x$ and  
$\varphi$,
the supersymmetry structure is that of two dimensions less.
This structure will continue in higher dimensions, namely in $2n+2$ dimensions
we will have $2n$-dimensional supersymmetry algebra for
the non-critical superstrings.

In order to construct the covariant hybrid formalism 
we need to work in a bigger superspace.
This can be seen, for instance, by noticing that we 
have in the RNS formalism four fermionic variables $(\psi, \psi_l)$ and 
$(b,c)$. In the hybrid formalism these four anticommuting fields 
have to be re-expressed in terms of four anticommuting target space new 
variables, namely 
two target superspace fermionic coordinates and their
conjugate momenta, for each of the left and right sectors.

In order to double the superspace, 
we can add one more supercharge 
\eqn\QA{
Q_{\dot +} = \oint
e^{ -\half \phi - {i\over 2} H + i x}\,,
}
which has the property that 
\eqn\comm{
\{Q_{+}, Q_{\dot +}\} = \oint e^{- \phi - i H}\,,
}
and therefore it is local w.r.t. $Q_{+}$. Note that the commutation relation 
between $Q_{+}$ and $Q_{\dot +}$ closes onto the translation generator 
of the longitudinal mode and Liouville directions
(see section 3). The new charge does not 
impose new constraints on the string spectrum.

Similarly, we introduce
\eqn\QAA{
Q_{\dot -} = \oint
e^{ -\half \phi +{i\over 2} H - i x}\,,
}
which has the same property 
\eqn\commn{
\{Q_{-}, Q_{\dot -}\} = \oint e^{- \phi + i H}\,.
} 
After the picture changing 
operation $Z$
\eqn\U{
Z =\{Q_B, \xi\} = 
2\p\phi b \eta e^{2 \phi} + e^{\phi} (G^{+} + G^{-}) + 2 b \p \eta e^{2\phi} + 
\p b \eta e^{2\phi} + c \p \xi\,,
} 
where $G^{\pm}$ are given in \scC\ with $n=0$,
is applied, we get on the RHS a translation generator, which 
involves the Liouville field $\varphi$ and the space coordinate $x$ (see next section). 
Using the two
charges  $Q_{+}, Q_{\dot +}$ (or  $Q_{-}, Q_{\dot -}$) 
we will construct a superspace with two fermionic coordinates  $\t^{+}$ and 
$\theta^{\dot +}$ (or $\t^{-}$ and 
$\theta^{\dot -}$)
and their conjugate momenta $p_{+},p_{\dot +}$
(or $p_{-},p_{\dot -}$).
We will follow an hybrid type formalism 
in order to construct the covariant description of the strings in
this superspace.

Consider  next $D=3$ ($n=1$).  
We have the bosonic fields $(x_1,x_2,x,\varphi)$. 
The fermions are $(\psi_{1}, \psi_2, \psi, \psi_{l})$ 
which are the fermion super-partners of the coordinates, and 
the ghost fields $(b,c, \eta,\xi)$. In the  
$-\half$ picture, we can construct eight candidates for the supercurrents
\eqn\supAaa{
e^{- {1\over 2}\phi + {i\over 2}(\pm H^{1} \pm H \pm \sqrt{3} x)}\,.
}
However, only two of them are mutually local w.r.t. each other
and close a supersymmetry algebra.
We can choose the supercharges $Q_{+,\alpha}, \alpha=1,2$ as
\eqn\supB{
Q_{+,1} =\oint e^{- {1\over 2}\phi +{i\over 2} (H^{1} - H -  \sqrt{3} x)}\,,~~~~~
Q_{+,2} = 
\oint e^{- {1\over 2}\phi + {i\over 2} (H^{1} + H +  \sqrt{3} x)}\,,~~~~~
}
which satisfy
\eqn\alg{
\{Q_{+,1}, Q_{+,2}\} = \int e^{-\phi + i H^1}\,,
}
and zero otherwise.
This is the translation operator at the picture $-1$, and 
 after the picture raising operation on the second charge, 
the anticommutator reads
\eqn\superchBB{
\{Q_{+,1}, Q_{+,2}\} = \oint \p y\,,
}
where $y=x_1+ i x_2$.

As before, there is another choice
of supersymmetry generators
\eqn\superchCC{
{Q}_{-,1} =
\oint e^{- {1\over 2}\phi + {i\over 2} (- H^{1} + 
H +  \sqrt{3}  x)}\,,~~~~~
{Q}_{-,2} = 
\oint e^{- {1\over 2}{\phi} +{i\over 2} (-{H}^{1} - {H}
 -  \sqrt{3}{x})}\,.
}
Adding the right moving sector we 
have two choices.
We can choose the same supercharges on the left and right sectors
$(Q_{+,\a},\bar Q_{+,\a})$ or
$(Q_{-,\a},\bar Q_{-,\a})$ to
have $(2,0)$ supersymmetry for the type IIB string, 
or we choose different sets
as
$(Q_{+,\a},\bar Q_{-,\a})$ or
$(Q_{-,\a},\bar Q_{+,\a})$ to
get $(1,1)$ 
 supersymmetry for the type IIA string.
We have $(Q_{+,\a})^{\dagger} =  Q_{-,\a}$.

In addition to the $U(1)$ R-symmetry \J\
we have the  bosonic $SO(2)$
acting as the Lorentz group on two coordinates ($x_1,x_2$).

Again, in order to construct the covariant hybrid formalism 
we will need to work in a bigger superspace.
We construct the supercharges $Q_{\dot +,\dot \alpha}, {\dot \alpha}=1,2$ as
\eqn\supdot{
Q_{\dot +,\dot 1} =
\oint e^{- {1\over 2}\phi + {i\over 2} (-H^{1} + H -  \sqrt{3} x)}\,,~~~~~
Q_{\dot +,\dot 2} = 
\oint e^{- {1\over 2}\phi +{i\over 2} (-H^{1} - H +  \sqrt{3} x)}\,.
}
We will use four superspace coordinates 
$\t^{\a}_{+}$ and $\t^{\a}_{\dot +}$ and their conjugate momenta.
Similarly, all the above can be repeated for 
 $Q_{-,\alpha}$ and   $Q_{\dot -,\dot \alpha}$
(see appendix).
 
In $D=5$ ($n=2$) some subtlety arises. In this case, 
one can construct four mutually local supercharges $Q^{a}_{+}= 
(Q_{+,\a}, Q_{+,\a'})$ with 
$\a,\a' =1,2$ from the left sector of the form 
\eqn\dcinqueA{
Q_{+,\a} = \oint e^{- \half \phi \mp {i\over 2}(H^{1} + H^{2}) 
- {i\over 2}(H + \sqrt{2} x)}\,, ~~~~
Q_{+,\a'} = \oint e^{- \half \phi \mp {i\over 2}(H^{1} - H^{2}) 
+ {i\over 2}(H + \sqrt{2} x)}\,.
}
It can be checked that they close on the 
translation generators (in the picture $-\half$) of the 
four dimensional space. In the same way 
there are four mutually local supercharges $\bar{Q}_{+,a}$ 
from the right sector.
This gives the $N=2$ supersymmetry of type IIB strings in $2n=4$ dimensions. 
In a similar way, for type IIA strings we use
$\bar{Q}_{-,a}$ 
from the right sector.

In order to construct the covariant hybrid formalism 
we need to work in a bigger superspace with six dimensional
$N=2$ supersymmetry structure.
As before we can add additional supercharges
$Q_{\dot +,a}$ ($a =1,\dots,4$) which are mutually local w.r.t. 
$Q_{+,a}$ (see appendix).
This, however, does not quiet work as before. 
The reason is that the original set of anticommuting 
fields is $(\psi_{l}, \psi, \psi_{i})$ with $i=1,\dots,4$ and $b,c$ 
(plus the fermionization of the superghosts)
are not enough to describe eight coordinates plus their 
conjugate momenta. Even adding the fermionization 
of the superghost, we have at most ten anticommuting variables to 
rewrite 16 anticommuting variables. 

Comparing to the study of critical superstrings compactified on
$CY_{2}$, the analysis performed in \BerkovitsIM\ can be repeated. In 
\BerkovitsIM\ only half of the supersymmetry is manifest and in a 
subsequent paper a formalism \BerkovitsDU\ 
with manifest $N=2$ supersymmetry of the target space is constructed by 
doubling the number of variables and imposing an harmonic constraint.
Besides the original 4 + 4 coordinates, one introduces 
4+4 anticommuting new coordinates and momenta which are not 
obtained from the original fermions and a constraint is implemented 
at the level of physical states. 
This will be discussed in section 4. 
The bosonic symmetry of the non-critical strings is 
$SO(4) \times U(1)_R$.

The last non-critical string is at $D=7$. In this  case, there 
are 16 supercharges, which are dependent. Again to fully 
realize the $\CN=(2,0)$ and $\CN=(1,1)$ supersymmetry, one 
has to add some auxiliary variables and constrain them by harmonic 
constraints. It seems that this example can be more easily 
treated using the pure spinor formulation
\lref\op{P.A. Grassi and Y. Oz, {\it to appear}.}
\op. 
 the bosonic symmetry of the non-critical strings is 
$SO(6) \times SU(2)_R$. Note that
the R-symmetry group is now $SU(2)_R$ since the
scalar $x$ is compactified
on a circle of self-dual radius, where there is an enhanced
$SU(2)$ affine symmetry  \MurthyES.


\subsec{Spectrum}

The BRST cohomology of the RNS non-critical strings has not been fully computed for
every $n$.
It has been computed for the case  $D=1$ ($n=0$) in
\refs{\ItohQB,\BouwknegtAM,\BouwknegtVA}.

{\it Two-dimensional strings: $D=1$, $n=0$}

The BRST cohomology consists of states at ghost numbers zero, one and two.
At ghost number one there are two types
of vertex operators. In the NS sector we have in the $-1$ picture
\eqn\T{
T_k = e^{-\phi +ik x + p_l \varphi} \ .
}
Locality with respect to the space-time supercharges
 $Q_{+}$ and $Q_{\dot +}$ 
 projects on half integer values of the momentum in the $x$-direction 
\eqn\m{
x: k \in Z +\half \ .
}
The introduction of a second supercharge 
 does not change the constraint on the spectrum.

In the Ramond sector we have in the $-\half$ picture the vertex operators
\eqn\R{
V_k=e^{-{\phi \over 2} +{i \over 2} \epsilon H+ ik x + p_l \varphi} \ ,
}
where $\epsilon = \pm 1$.
Locality with respect to the space-time supercharges
 $Q_{+}$ and $Q_{\dot +}$ \Q.
implies $k \in Z + \half$ for $\epsilon=1$ and
$k \in Z$ for $\epsilon=-1$.
An interesting operator that we will consider later is
the supercurrent
\eqn\Ramond{
V_{k=1,\epsilon=-1}=e^{-{\phi \over 2} -{i \over 2} H+ ix}\,.
}

The Liouville dressing is determined by
requiring conformal invariance of the integrated vertex operators. Thus,
the coefficient $p_{l}$ has to be a solution to the equation
\eqn\bet{
{k^2 \over 2}-\half p_l\left(p_l-2\right)=\half \ .
}
This equation can be solved by $p_{l} = 1\pm k$. 
Furthermore the locality constraint requires $p_{l} \le {Q \over 2}= 1$.
Being in the BRST cohomology imposes an additional constraint
in the Ramond sector $|k| = -\epsilon k$ \ItaNE.

Note that 
\eqn\dag{
(T_k)^{\dagger} = T_{-k},~~~~(V_{k,\varepsilon=\pm 1})^{\dagger}
= V_{-k,\varepsilon=\mp 1}\,.
}
At ghost number zero 
there are spin zero BRST invariant operators that generate a commutative, associative
ring
\eqn\B{
{\cal O}(z){\cal O}'(0)\sim {\cal O}''(0)+\{Q_B,\dots\} \ ,
}
called the ground ring, where $Q_B$ is the worldsheet $N=1$ BRST operator.

The main objects in the construction of the ring are
the R-sector operators
\eqn\chiralring{
x(z)=
\left( e^{-{i\over 2}H}e^{-{1\over 2}\phi}
-{1\over\sqrt{2}}e^{{i\over 2}H}\partial\xi e^{-{3\over 2}\phi}c
\right) e^{{i\over 2}x-{1\over 2}\varphi} \ , 
}
$$
y(z)=
\left( 
e^{{i\over 2}H} e^{-{1\over 2}\phi}
-{1 \over \sqrt{2}}e^{-{i\over 2}H}\partial\xi e^{-{3\over 2}\phi}c
\right) e^{-{i\over 2}x-{1\over 2}\varphi} \ ,
$$
and the NS-sector operators 
\eqn\NS{
u=x^2,~~~ 
v=y^2,~~~ 
w=xy \ .
}
Locality with respect to the space-time supercharges
 $Q_{+}$ and $Q_{\dot +}$ implies the projection
$x \rightarrow -x$. Thus, the basic invariant elements are
$x^2$ and $y$. We have $x^\dagger = y$. 

The ghost number two operators correspond to spin one currents, 
 that acts as derivations of the ring.

In order to construct the
string states we need to combine the left and right moving states in such
a way that the momentum along the Liouville direction, which is non-compact,
is the same in both sectors. 
Projecting on the left and the right sectors with the same
set of supercharges \Q\ 
or \QA\ defines the type IIB theory.
Projection
on the left and the right sectors with different set
of supercharges 
defines the type IIA theory \refs{\MurthyES\SeibergBX\ItaNE}.


{\it Higher dimensions}

As noted above, the complete BRST cohomology has not been
computed yet.
We consider, for instance  at ghost number one, 
two types
of vertex operators similar to \T\ and \R.
In the NS sector we have 
\eqn\Th{
T_k = e^{-\phi +ikX+p_l \varphi}V(z) \ ,
}
where $V(z)$ is an $N=1$ primary made of the $2n$ superfields
$X^i$.
One has  
\eqn\dim{
\Delta + {k^2 \over 2}-\half p_l\left(p_l-Q\right)=\half \ ,
}
where $\Delta$ is the dimension of $V$.
If  $V$ has $U(1)$ charge $q$ 
then 
locality with respect to the supercharges implies that
\eqn\met{
k Q +q \in  2 Z + 1 \,. 
}
When $Q=2$ and $q=0$ we recover \m.  

In the Ramond sector consider, for instance, the vertex operators
\eqn\Rh{
V_k=e^{-{\phi \over 2} +{i \over 2} \epsilon H
+ {i \over 2}\sum_{I=1}^n  \epsilon_I H^I+ ikX + p_l \varphi}\,,
}
where $\epsilon,\epsilon_I= \pm 1$.
Locality with respect to  the supercharges implies  
\eqn\cond{
kQ \in 2Z -\half \left(\epsilon + \sum_{I=1}^n \epsilon_I- 1\right)\,,
}
and
\eqn\condition{
kQ \in 2Z -\half \left(\epsilon + \sum_{I=1}^n \epsilon_I- 1\right)\,.
}
Again for $Q=2$ we recover the two-dimensional results.
These conditions can be solved. 
As an example consider the case $D=3$ ($n=1$).
Then we have 
\eqn\conditionone{
kQ \in  2Z -{1\over 2}\,,
}
when $\epsilon=\epsilon_1=\pm 1$ and
\eqn\conditiontwo{
kQ \in  2Z  + {1\over 2}\,,  
}
when $\epsilon=-\epsilon_1=\pm 1$.
Interesting operators that we will consider later are
the supercurrents
\eqn\Ramondnext{
V_{k=\pm{\sqrt{3} \over 2},\epsilon_1=-1,\epsilon=\mp 1}=
e^{ - {1\over 2}\phi - {i\over 2} H_{1}  
\mp {i\over 2} (H -   \sqrt{3}) x}\,. 
}

\newsec{Covariant Non-Critical Superstrings}
\subsec{$D=1$, $n=0$}

In the following we define superspace variables,
which exhibit a similar structure to 
that of \BerkovitsTG, where compactification on a Calabi-Yau 4-fold is discussed.
Let us discuss the left-moving sector, 
and everything should be replicated for the right sector.
However, as discussed before, 
when we combine the left and right sectors, there is a choice 
corresponding to the type IIA and type IIB GSO projections
in the RNS formalism. 
As already explained before we use two supercharges $Q_{+}$ and 
$Q_{\dot +}$ 
to construct the covariant formalism.

Consider the two supercharges \Q\ and \QA.
In order to define the supersymmetry off-shell, we change pictures
and modify $Q_{+}$ into $Z Q_{+}$ as
\eqn\qq{
Q_{+}  = \oint \left(b \eta e^{ {3\over 2} \phi - {i\over 2} H - i x}
+ {1\over 2}(\p \varphi - i \p x + 2 \p \phi)e^{ {1\over 2} \phi + {i\over 2} H - i x} 
- e^{{1\over 2} \phi - {3 i\over 2} H - i x}
\right) \ .
}
The first term is the application of the picture changing operator (PCO)
\U\ on $Q_{+}$ in the $-\half$ picture. The 
additional terms and the $\phi$ dependence is coming from 
the non-homogeneous term in \scC.\foot{A way to check that this gives 
the correct answer is to apply the PCO to $Q_{\dot +}$ and on the r.h.s. of 
the \comm\ to check that they give the same answer.}
Then $Q_{+}$ and $Q_{\dot +}$ satisfy

\eqn\qqA{
\{Q_{+}, Q_{\dot +}\} = \oint \p(\varphi - i x + 2 \phi)\,,
}
where we notice that the translation operator contains in addition
to $x$, 
both the Liouville field $\varphi$ and the superghost $\phi$. 

We construct superspace variables as the dimension zero combinations 
\eqn\suA{
\theta^{+}  = c \xi e^{- {3\over 2} \phi + {i\over 2} H + i x}\,, ~~~~~
\theta^{\dot +} = 
e^{ \half \phi + {i\over 2} H - i x}\,.
}
The variables $\theta^{+}$ and $\theta^{\dot +}$ have regular OPE, and 
\eqn\susy{
q_{+}(z)\theta^{+}(w)  \sim  {1 \over (z-w)},~~~~~~
q_{\dot +}(z)\theta^{\dot +}(w)  \sim  {1 \over (z-w)}\,.
}
The conjugate 
momenta  to $\theta^{+}$ and $\theta^{\dot +}$ are the dimension one objects
\eqn\shB{
p_{+} = b \eta e^{{3\over 2} \phi - {i \over 2} H  - i x}\,, ~~~~
p_{\dot +} = e^{ -  \half \phi - {i\over 2} H + i x}
}
and 
\eqn\pth{
p_{+}(z)\theta^{+}(w) \sim  {1 \over (z-w)},~~~~~~
p_{\dot +}(z)\theta^{\dot +}(w) \sim  {1 \over (z-w)} \ .
}
Notice that we defined the conjugate 
momenta with the a different sign for the $x$ part, which does not 
change the conformal spin.
With this choice $p_{+}$ and $p_{\dot +}$ have regular  OPE.

Let us discuss the 
hermiticity properties and the periodicity 
of $x$. As we discussed before in the RNS formalism,
 $Q_{+}^{\dagger} = Q_{ -}$ and $Q_{\dot +}^{\dagger}= Q_{\dot -}$.
However, in order to have manifest space-time supersymmetry,
we had to apply the picture changing operation. With this, the definition
of the hermiticity conditions gets more complicated  \BerkovitsBF.
Let us define now the compatible hermiticity conditions.

We define the  
hermiticity conditions by 
\eqn\hermA{
(x_{m})^{\dagger } = e^{R} x_{m} e^{-R}\,,~~~~~
(\varphi)^{\dagger } = e^{R} \varphi e^{-R}\,,~~~~
(\psi_{m})^{\dagger} = e^{R} \psi_{m} e^{-R}\,, 
}
$$
(e^{\phi})^{\dagger} = e^{R} e^{\phi + \Delta \phi} e^{-R}\,, ~~~~
(e^{\chi})^{\dagger} = e^{R} e^{\chi + \Delta \chi} e^{-R}\,, ~~~~
(e^{\kappa})^{\dagger} = e^{R} e^{\kappa + \Delta \kappa} e^{-R}\,, ~~~~
 $$
where $R$ is given by (see appendix (7.2))
\eqn\Rt{
R = \oint \Big[
(G^{+} + G^{-}) e^{\chi - \phi - \kappa} + {1\over 2} \p\phi 
e^{2 (\chi - \phi - \kappa)} \Big]\,,
}
and $G^{\pm}$ are the supersymmetry generators for the matter system 
given in \scC.
The shifts $\Delta$ in the ghost fields are given by
\eqn\hermB{
\Delta \phi = 2 \chi - 2 \kappa - 4 \phi\,,~~~~
\Delta \chi = - 2 \kappa - 2 \phi\,, ~~~~
\Delta \kappa = - 2 \chi + 2 \phi\,. 
}
The importance of these shifts 
will become apparent shortly. 
Note that in addition one applies in the hermiticity definition
ordinary complex conjugation.

With this definition of hermiticity we can check that 
 $Q_{+}^{\dagger} = Q_{ -}$. This can be seen 
by observing that 
the supersymmetry charge $Q_{+}$  can be rewritten 
\eqn\simQ{
Q_{+} = e^{R} 
\Big(\oint b \eta e^{{3\over 2} \phi - {i\over 2} H - i x}\Big) e^{-R}\,,
}
and that with the definitions \hermA\ and  \hermB 
\eqn\qd{
Q_-^{\dagger} = \oint b \eta e^{{3\over 2} \phi - {i\over 2} H - i x}\,.
}
Also,
\eqn\td{
\t_{+}^{\dagger} = \t_{-},~~~~
p_{+}^{\dagger} = p_{-}\,.
}
Note that since $R^{\dagger} =R$, we have $(O^{\dagger})^{\dagger}=O$,
as required.

The hermitian conjugation does not commute with the stress energy-momentum 
tensors. This implies that the conformal weights of some fields 
and their hermitian conjugates are different. In addition, we observe that 
there is a combination of the ghost fields $\phi, \chi, \kappa$, namely 
$ \phi +\kappa - \chi$, which is invariant under the shifts \hermB. 
This combination appears 
in the definition of $R$.  

The same hermiticity conditions will be used in higher dimensions.
The only difference is that in the definition
of $R$ \Rt\ 
we will need the appropriate $G^{\pm}$.

The fermionic fields $\t_{+}, \t_{\dot +}, p_{+}, p_{\dot +}$ 
have singular OPE's with the field $x$. A way to 
solve this problem is to redefine the variable $x$ such that 
there are no singular OPE's by performing a 
similarity transformation on the operators $q_{+}$ and $q_{\dot +}$ 
and the translation generator. Also the energy-momentum tensor 
$T$ is modified. 
The similarity transformation is given 
by 
\eqn\simA{
U = exp \oint (i x + i H - \varphi - \phi) \p \kappa \,. 
}
The combination $ (i x + i H - \varphi - \phi)$ has several nice properties: 
it has no singularities with itself, it has no singularities with the Grassman 
variables $\theta^{+}, \theta^{\dot +}, p_{+}, p_{\dot +}$ (but it has singularities with the supercurrents), and it shifts the translation operator in the r.h.s. of 
\qqA\ by $ 2 i \oint \p \kappa$. 
It is therefore convenient to introduce the new coordinate $x'$ 
defined by 
\eqn\alA{
x' = x + 2 i (\phi + \kappa)\,,
}
such that \qqA\ becomes 
\eqn\QQ{
\{Q_{+}, Q_{\dot +}\} = \oint \p(\varphi - i x')\,.
} 
It can be easily verified that 
\eqn\xp{
x'(z) x'(w)  = - \ln(z-w)\,,
}
since the 
contributions of $\phi$ and $\kappa$ cancels. The
 primary field $e^{a x'}$, where $a$ is complex, 
transforms under the hermiticity condition as follows
\eqn\hermC{
(e^{a x'})^{\dagger} = e^{R} (e^{\bar a x'}) e^{-R}\,,
}
where we use 
$\phi + \kappa + \Delta (\phi + \kappa) = - (\phi + \kappa)$. 
So, the new variable $x'$ is (modulo the similarity 
transformation $e^{R}$) hermitian. 

We stress again that as in the RNS description  where the
hermitian  conjugation
relates the model to its conjugate (with 
the GSO projection being performed by the charges $Q_{-}$ and 
$Q_{\dot -}$), the same holds here. 
In particular,
the supersymmetry 
algebra \QQ\ is mapped into the conjugated relation
\eqn\QQc{
\{Q_{-}, Q_{\dot -}\} = \oint \p(\varphi + i x')\,.
} 
Obviously, the spectrum of the theory and its conjugate
are equivalent as 
can be easily observed by mapping the observables using the 
hermitian conjugation.

As we stressed before,  
in order to have space-time
supersymmetry the field $x$ is taken to be
periodic of period $Q^{-1}(n)$.
One may worry whether the field redefinition or the hermitian conjugation
interferes 
with periodicity. 
However, this is not the case
since the field redefinition is obtained by acting with $U$ given in \simA, 
which is 
invariant under constant translations of $x$. The periodicity 
of $x'$ (which is the same as of $x$) is also not affected
by the hermitian 
conjugation.  
In the RNS framework, the periodicity of
$x$
is needed in order to ensure that the interacting theory
is space-time supersymmetric.
The worldsheet operator that changes the radius $\int \p x \bar{\p}x$ is not
$N=2$ invariant, and when switched on it breaks the 
worldsheet $N=2$ superconformal algebra leading to the breaking of
space-time supersymmetry.
Similarly, one can argue the same using the 
supersymmetric variables.

Note also that the current 
\eqn\alB{
 J_{x'} = \p x' = \p x + 2 i (\p \phi + \p \kappa)} 
has no singularity with $\t^{+}$ and $\t^{\dot +}$. 
The combination \alA\ will appear again in the higher dimensional cases
in the form
\eqn\alc{
x' = x + i Q(n) (\phi + \kappa)\,.
}

In the next step we rewrite the ghost fields in terms of new chiral 
bosons $\omega$ and $\rho$ by imposing the following two equations
\eqn\cbA{
b = p_{+} \e^{\omega - \rho} \,, ~~~~~
- \gamma^{2} b = p_{\dot +} e^{\omega + \rho}\,.
}
 The conformal spins of the combinations $e^{\omega -\rho}$ and 
 $e^{\omega + \rho}$ are 1 and 0, respectively. 
These conditions lead to the following equations 
\eqn\cbB{
\kappa  + {3\over 2} \phi - {i \over 2} H - i x  + \omega - \rho = 0\,, ~~~
}
$$
- 2(\phi + \kappa)  + \chi  - \half \phi  - {i \over 2} H + i x + \omega + \rho
=0\,. 
$$
From this we get the solution for $\omega$ and $\rho$
\eqn\cbC{
\omega = \half(\phi + \kappa - \chi + i H)\,, ~~~~~
\rho = 2 \phi  + \half(3 \kappa - \chi) - i x = 
- {1\over 2}\kappa - \half \chi - i x'\,.
}
The hermiticity properties of these new fields can be 
deduced from the transformation laws of the original 
fields \hermA\
\eqn\herrho{
(e^{\rho})^{\dagger} = e^{R} e^{\rho + \Delta \rho} e^{-R}\,, ~~~~
(e^{\omega})^{\dagger} = e^{R} e^{\omega} e^{-R}\,,
}
where $\Delta \rho = - 2 \rho$. 
Under the hermitian conjugation defined in \hermB\ the chiral bosons 
$\omega$ and $\rho$ are mapped into the corresponding chiral 
boson of the conjugate theory. This can be deduced by constructing 
the conjugate theory and the corresponding ghosts, 
from the set of charges: $Q_{-}$ and $Q_{\dot -}$. 

It is easy to check using the above definitions
that the OPE's of these new fields are 
\eqn\cbD{
\rho(z) \rho(w) \sim - \half \log(z-w)\,, ~~~~
\omega(z) \omega(w) \sim \half \log(z-w)\,, ~~~
\rho(z) \omega(w) \sim 0\,.
}
The
stress energy tensor reads
\eqn\cbE{
T_{\omega,\rho} = (\p \omega)^{2} - \p^{2} \omega  - (\p \rho)^{2} - \p^{2} \rho\,.
}
Therefore, the conformal spin of $e^{\omega}$ is $3/4$, $e^{\rho}$ is $-3/4$ and 
$e^{-\rho}$ is $1/4$ which is consistent with $e^{\omega+\rho}$ which 
has conformal spin 0 and $e^{\omega-\rho}$ with conformal spin 1.\foot{The OPE 
of $T$ with $e^{\a \omega}$ and $e^{\b\rho}$ are  
$
T_{\omega,\rho} e^{\a \omega} = \Big( {\a(\a/2 +1)/2 \over (z-w)^{2}} + {\p \over (z-w)} \Big) e^{\a\omega}\, 
$
and
$
T_{\omega,\rho} e^{\b \rho} = 
\Big( {-\b(\b/2 +1)/2 \over (z-w)^{2}} + {\p \over (z-w)} \Big) e^{\b\rho}
$.
} 

Notice that  
the combination 
\eqn\Jp{
J^{\prime} = - 2 \p \rho - 2 i J_{x'}
}
reads 
\eqn\j{
J^{\prime} = - 2 \p \Big( 2\phi + {3\over 2} \kappa - {1\over 2} \chi - i x\Big) 
- 2 i \p \Big( x + 2 i \phi + 2 i \kappa \Big) = \p \kappa + \p \chi = cb + \eta \xi\,,
}
as in \scA\ where we used the 
definitions \cbC.

Using these new variables, we write the energy momentum tensor as
\eqn\alB{
T^{\prime} = - p_{+} \p \t^{+} - p_{\dot +} \p \t^{\dot +} + (\p \omega)^{2} - \p^{2} \omega 
- (\p\rho)^{2} - \p^{2} \rho 
}
$$
- \half (\p x')^{2} - i \p^{2} x' - {1\over 2} 
(\p \varphi)^{2} + \p^{2} \varphi \,.
$$
{\it To summarize}: we replaced the  four bosonic 
variables $(x, \varphi, \beta, \gamma)$ 
and four fermionic 
variables $(\psi,\psi_{l}, b,c)$ in the RNS formulation by
 four bosonic  variables  $(x',\varphi,\omega, \rho)$ 
and four fermionic
 variables  $(p_{+}, \theta^{+}, p_{\dot +},\theta^{\dot +})$.
Let us
now compute the total central charge to check the consistency 
of the above manipulations.
We have the following contributions 
$$(-2)_{p_{\dot +} \t^{\dot +}}+ (-2)_{p_{+} \t^{+}}+
(1-6)_{\omega}+ (1+6)_{\rho}+(1-12)_{x'}+ (1+12)_{\varphi}=0\,.$$

Now we have to look for an $N=2$ superconformal algebra written 
in terms of these new variables. 
The generators $T^{\prime}, J'$ are the generators that appear in the twisted
$N=2$ superconformal algebra \scA\ written in the new variables.
The other $N=2$ generators are given by the supersymmetry  charge
\eqn\alD{
G^{\prime -} = p_{+} e^{\omega -\rho}\,, ~~~~
}
which has conformal spin $2$ and ghost number $-1$ w.r.t. $J'$,  
and the BRST charge $G^{\prime +}$ (see appendix)  
\eqn\BRSTA{
G^{\prime +} =
\left(p_{\dot +} +  \t^{+} (\p \varphi - i \p x)\right)
e^{\rho + \omega} +
p_{\dot +} \left[
(\p \varphi + i \p x') - 2 (\t^{+} p_{+} + \p \omega +\p \rho) \right]
e^{\rho - \omega}
}
$$
+\t^{+} e^{\rho - \omega} T' + \p(\t^{+} e^{\rho -\omega} J') +
\p^{2}(\t^{+} e^{\rho -\omega})\,.
$$

There is another set of supersymmetric
invariant variables that will be useful when considering
curved target spaces with RR background fields.
It is given by
 $$
 \Pi_{+\dot +} = \p (\varphi - i x')\,, ~~~~
 \Pi^{+ \dot +} = \p (\varphi + i x') - 2 \t^{+} \p\t^{\dot +}\,, ~~~~ 
 $$
 \eqn\newVA{
 d_{+} = p_{+}\,, ~~~~~ 
 d_{\dot +} = p_{\dot +} + \t^{+} \Pi_{+\dot +}\,, 
 }
 which satisfy the algebra
\eqn\newVB{
d_{+}(z) d_{\dot +}(w) \sim {\Pi_{+ \dot +} \over (z-w)}\,, ~~~~
\Pi^{+\dot +}(z) \Pi_{+\dot +}(w) \sim {-2 \over (z-w)^{2}}\,, 
}    
$$
d_{+}(z) \Pi_{+\dot +}(w) \sim 0\,, ~~~~
d_{\dot +}(z) \Pi_{+\dot +}(w) \sim 0\,, ~~~~
$$
$$
d_{+}(z) \Pi^{+\dot +}(w) \sim - 2 {\p \t^{\dot +} \over (z-w)}\,, ~~~~
d_{\dot +}(z) \Pi^{+\dot +}(w) \sim  - 2 {\p \t^{+} \over (z-w)} \,. 
$$
We will use the notation
 $({}_{+\dot +},{}^{+\dot +}, +, \dot +)$ for the superspace indices
\eqn\xpp{
x_{+ \dot +} = \varphi - i x', 
x^{+ \dot +} = \varphi + i x', \t^{+}, \t^{\dot +}\,.
} 
The algebra of fermionic derivatives 
\eqn\derivatives{
D_{+} = \p_{\t^{+}}\,,~~~D_{\dot +} = \p_{\t^{\dot +}} + \t^{+} \p_{x_{+\dot +}}\,,
} 
is given by 
\eqn\alE{
\{D_{+}, D_{+} \}=0\,, ~~~~
\{D_{+}, D_{\dot +} \}=\oint \Pi_{+\dot +} = \p_{x_{+\dot +}} \,, ~~~~
\{D_{\dot +}, D_{\dot +} \}=0\,. ~~~~
}
The supercurrents commute with the covariant derivatives and 
therefore it is easy to see that they have the form 
\eqn\alEA{
q_{+} = p_{+} - \t^{\dot +} \Pi_{+\dot +}\,, ~~~~ 
q_{\dot +} = p_{\dot +}\,.
} 
The algebra of the supercurrents is very similar to the algebra of 
covariant derivatives except the sign in front of the translation 
operator appearing in the OPE $q_{+}(z) q_{\dot +}(w)$.
In terms of the new variables, the energy-momentum tensor 
can be recast as
\eqn\newT{
T' = - d_{\dot +} \p \t^{\dot +} - d_{+} \p \t^{+} - {1\over 2} \Pi_{+\dot+} \Pi^{+\dot +} 
+ {Q\over 2} \p \Pi_{+\dot +} 
}
$$
+ (\p \omega)^{2} - \p^{2} \omega - (\p\rho)^{2} - \p^{2}\rho\,.
$$

At this point we are done with the change of variables.
However, it is of value to  
to uncover now an algebraic relation between the $2n+2$ 
non-critical superstrings
and the critical superstrings compactified on $CY_{4-n}$ manifolds.
In the $n=0$
two-dimensional case we compare to \BerkovitsTG.
The form of the energy momentum tensor \newT\
is identical to that given in \BerkovitsTG
(see eq.~(2.11) there), except for the fact that there is the 
background charge associated with $x'$ and $\varphi$ and the term 
$T_{GS}$ is missing. 
Let us construct now a twisted  $N=2$ algebra with $\hat{c}=-2$ which will be compared
with a similar algebra in  \BerkovitsTG.
For that we 
choose as the $N=2$ 
$U(1)$ current the operator $J'$  obtained by adding the contribution of the Liouville
$$
J' \rightarrow J' + 2 \p \varphi = - 2 \p \rho + 2 \p(\varphi - i x')\,.
$$
Now with the modified $U(1)$ current we have
\eqn\newAN{
J'(z) J'(w) \sim {- 2 \over (z-w)^{2}}\,, ~~~~
T'(z) J'(w) \sim { 2 \over (z-w)^{3}} \,.
}
Notice that in computing $T'(z) J'(w)$ the combination 
$\varphi - i x'$ has no singularities with term $Q \Pi_{+\dot +}$ in $T'$. 
Next we perform a similarity transformation by 
$T' \rightarrow e^{R} T' e^{-R}$ and $J' \rightarrow e^{R} T' e^{-R}$
(not to be confused with $R$ of \Rt) 
with 
\eqn\R{
R = 2 \oint (\varphi - i x') \p \rho \,.
}
This removes the 
terms 
$Q \Pi_{+\dot +}$ and $2 \p(\varphi - i x')$ from $T'$ and from $J'$. 
Then, the form of $T'$ and $J'$ are
\eqn\TJ{
T' = - d_{\dot +} \p \t^{\dot +} - d_{+} \p \t^{+} - {1\over 2} \Pi_{+\dot+} \Pi^{+\dot +} 
}
$$
+ (\p \omega)^{2} - \p^{2} \omega - (\p\rho)^{2} - \p^{2}\rho\,.
$$
$$
J'  = - 2 \p \rho\,,
$$
and the rest of the algebra is given by 
\eqn\algAA{
G^{-'} = d_{+} e^{\omega-\rho}\,, ~~~~~
G^{+\prime} = e^{-\oint \Pi_{+\dot +} e^{- 2 w}} d_{\dot +} 
e^{\rho -\omega} e^{\oint \Pi_{+\dot +} e^{- 2 w}}\,.
}
This is exactly the algebra  
in \BerkovitsTG\ 
where the terms $T_{GS}, G^{+}_{GS}, G^{-}_{GS}$ and $J_{GS}$ are 
set to zero.

Let us comment on the relation between these algebras.
In the case studied in  \BerkovitsTG\ the string theory is compactified 
on a Calabi-Yau fourfold $CY_{4}$.
The CFT on   $CY_{4}$  has central charge
 $\hat c = c/3 = 4$. The total charge of string theory is the sum of the CFT on $CY_{4}$ and 
 the central charge of the uncompactified part which is $\hat c = c/3 = -2$,
which is the same central charge of the algebra \TJ\ and \algAA.
The sum
 gives $\hat c = 2$ which is indeed the central charge of the RNS algebra
\scA.

This structure continues to hold in higher dimensions as we will see. 
We compare the non-critical strings for 
 $D= 2 n + 1, n =0,1,2,3$ with the string theory compactified on $CY_{4-n}$. 
The 
uncompactified sector has central charge $\hat c_{un} = 2 - (4 - n) =  n-2$.
So, the central term of the appropriate $J'$ in the algebra
for the non-critical string in $D=1 +2 n$ should be 
 $(n-2)/(z-w)^{2}$. In the case $n=0$ we have exactly $ -2 /(z-w)^{2}$ as in \newAN. 

We point out that in the new formulation the $N=2$ superconformal algebra 
plays a fundamental role. Indeed it is used to characterize the physical 
(supersymmetric) states of the theory. In the original formulation the 
spectrum is characterized by  the BRST cohomology in the small Hilbert 
space (without the zero mode of $\xi$). The BRST cohomology can be also
 computed in the large Hilbert space (with the zero mode of $\xi$) by selecting 
a finite number of pictures. In the new framework (which is built on the 
large Hilbert space) the BRST condition on vertex 
operators is replaced by the condition of being chiral primary w.r.t. 
the $N=2$ superconformal algebra. The selection of a finite number of pictures 
is obtained by selecting a finite number of power of $e^{\rho}$.
 In \BerkovitsBF, 
the equivalence of the two descriptions is exploited. 


\subsec{The two-dimensional spectrum in the new variables}

Let us compute the spectrum in the new variables. 
It is convenient to find the inverse map between the original variables and 
the new supersymmetric variables. Bosonizing the 
fermions $\t^{+}, \t^{\dot +}, p_{+}$ and $p_{\dot +}$ 
by 
\eqn\fer{
\t^{+} = e^{\alpha}, p_{+} = e^{- \alpha}, 
\t^{\dot +}= e^{\beta}, p_{\dot +} = e^{-\beta}\,,
}
we have the relations
\eqn\reA{
i H =  {1 \over 2} (\alpha + \beta + 2 \omega) \,,~~~~~~~
\phi = - {1\over 2} (- 3 \alpha + \beta + 2 \omega - 8 \rho - 4 i x')\,,
}
$$
\chi = \alpha - \omega + \rho\,,~~~~~~~~~~~
\kappa = -\alpha + \omega - 3 \rho  - 2 i x'\,.
$$

In the NS sector , see \T, we have 
\eqn\TA{
T_{k} = e^{\Big(\alpha \left(-{3\over 2}+k\right) +
\beta \left(\half - k\right) + 
\omega + (-2 + k) (2\rho + i x') + p_l \varphi
\Big)}\,, ~~~~~ k \in Z + \half\,,
}
and in the R sector, see \R\ for $\epsilon = \pm$, we have 
\eqn\RA{
V_{k,\epsilon=+1} = e^{\Big(
\beta\,\left( {1 \over 2} - k \right)  + 
  \alpha\,\left( - {1 \over 2}   + k \right)
      + \omega + \left( -1 + k \right) \,
   \left( 2\,\rho + i x' \right) + p_l \varphi\Big)}\,, ~~~~
   k \in Z + \half\,,
}
$$
V_{k,\epsilon=-1} = e^{\Big(
\alpha\,\left( -1 + k \right)  - \beta\,k + 
  \left( -1 + k \right) \,
   \left( 2\,\rho + i \, x' \right)
   + p_{l} \varphi
   \Big)}\,, ~~~~~ k \in Z\,.
$$
We notice that in both cases, the powers of $\omega$ and 
$\rho$ have only integer values. The value of $p_l$ 
is fixed by the conformal 
dimension of the integrated vertex operator. 
Note that 
$e^{k\alpha}$ and $e^{k\beta}$ have conformal weight  $k(k-1)/2$,
and  $e^{kx'}$  has conformal weight  $k(k+2)/2$.
One can easily check that 
$T_{k}$ 
and $V_{k}$ have the correct dimensions. 

From equations \TA\ we immediately see that, in the 
case of NS vertex operators, the momentum 
$k$ must be half integer in order to rewrite it in terms 
of the new variables, and for the R sector we have that for $\epsilon = +1$ 
$k$ must be half integer, while for $\epsilon = -1$, the momentum should 
be an integer. This is the way that the locality with respect 
to the space-time supercharges in the RNS formalism is realized
in the hybrid variables. 

In order to express the above vertices in terms of the 
new variables, we need a dictionary. It is easy to check that 
\eqn\diA{
e^{\a} = \t^{+}\,, ~~~
e^{2 \a} = \p \t^{+} \t^{+}\,, ~~~
e^{3 \a} = \p^{2} \t^{+} \p \t^{+} \t^{+}\,, \dots\,,
}
$$
e^{-\a} = p_{+}\,, ~~~
e^{-2\a} = \p p_{+} p_{+}\,, ~~~
e^{-3 \a} = \p^{2} p_{+} \p p_{+} p_{+}\,, \dots\,.
$$ 
The order of the expression can be easily determined by 
observing that 
$e^{2 \a} = \lim_{x\rightarrow y} :\p\t^{+}(x) \t^{+}(y): = 
\lim_{x\rightarrow y}  :\p \a e^{\a}(x) e^{\a}(y):$ 
Using the OPE $e^{\a}(x) e^{\a}(y)  \rightarrow (x-y) e^{2\a}(y)$, 
then it is easy to check the above formulas. 
In addition, one can check the OPE's between different monomials.  

Let us examine the Liouville-independent states. 
By setting $p_l = 0$, 
we have the state \Ramond\
\eqn\tA{
V_{+1,\epsilon=-1} = p_{\dot +}\,.
}
It has a very simple interpretation using 
the definition of the supercurrent 
$q_{\dot +} = p_{\dot +}$.
The vertex operator $V_{+1,\epsilon=-1}$ describes the single fermion 
of the open string theory. It is massless and it does not depend on the 
coordinate $x'$. The coupling  
can be done by adding the deformation 
\eqn\intv{
S_{\psi} = t_{R} \oint d \sigma \psi  q_{\dot +}\,,
} 
where $\psi$ is a 
constant Grassmann number. 
Note in comparison that in \BerkovitsTG\
there are two massless fermions which are compactification 
independent states. 

Combining left and right sectors, we have for constant RR fields 
of IIB/A string theories the vertex operators 
\eqn\tB{
V^{A}_{RR} = q_{\dot +} \bar q_{\dot -}\,, ~~~~~
V^{B}_{RR} = q_{\dot +} \bar q_{\dot +}\,, 
} 
where $\bar q_{\dot +}$ and $\bar q_{\dot -}$ are the right moving 
charges. Again, the number of independent RR vertex operators is 
dictated by by the level matching, and the BRST invariance. 
As will be explained in section 5, the coupling of the RR vertex operators
to the space-time RR fields strength $F^{\alpha\beta}$ is
\eqn\RRc{
F^{{\dot +}{\dot -}}q_{\dot +} \bar q_{\dot -}\,, ~~~~~
F^{{\dot +}{\dot +}}q_{\dot +} \bar q_{\dot +}\,.
}
Thus we find one 
RR scalar both for IIA and IIB.
In Type IIB this corresponds to a self-dual 1-form field strength in two dimensions.
In Type IIA this corresponds to a 2-form field strength, or alternatively, its scalar
Hodge dual.  

As discussed above,  in the RNS description
there are ghost number zero dimension zero states that
generate the ground ring.
Let
us write them in the new variables. 
The ghost number zero dimension zero operators are given by 
\eqn\gA{
x(z) =\Big( e^{- \half \a - \half \b} - 
{\p \kappa
\over \sqrt{2}} e^{\half \a +\half \b}\Big) e^{-  \rho - {i \over 2} x' - \half \varphi}\,,
}
$$
y(z)  =\Big(
e^{\omega} -
{\p \kappa
\over \sqrt{2}} e^{-\omega} \Big)
e^{-\a + \b -  3 \rho - {3i \over 2} x' - \half \varphi}\,.
$$ 
We see that $x(z)$ cannot be written in terms of the
variable \diA, which corresponds to the fact that it has been projected
out by the GSO projection in the RNS formalism.
$y(z)$ takes the from
\eqn\Y{
y(z) = 
\Big(p_{+}\t^{\dot +} - {e^{-2\omega} \over \sqrt{2}}
\Big[\p p_{+}\t^{\dot +} + p_{+}\t^{+}(\p \omega -3\p \rho - 2i \p x')\Big]\Big)
e^{\omega - 3 \rho - {3i \over 2} x' - \half \varphi} \,.
}
 Notice that compared with \ItaNE\ the 
surviving state is $y$ and not $x$ because we project here 
with respect to the $q_{+}$ supercurrent.

The operators $u=x^2,v=y^2$, which are not projected out by the GSO projection
in the RNS formalism  \ItaNE\
take the form
$\t^{+}, \t^{\dot +}, p_{+}$ and $p_{\dot +}$
\eqn\newUV{
u(z) = - p_{+} \p p_{+} \t^{\dot +} \p \t^{\dot +} e^{\omega 
- 3 \rho - \half \varphi - {3 i/2} x'} + \dots\,, ~~~~~~~     
}
$$  
v(z) = p_{+} p_{\dot +} e^{- 2 \rho - \varphi - i x'} + \dots \,,
$$
while $w=xy$ which is projected out cannot be written in the supersymmetric variables.
One can now repeat the analysis of \ItaNE\ in a manifestly space-time
supersymmetric manner.


\newsec{Higher Dimensions}

\subsec{$D=3$, $n=1$}

In this case the bosonic fields  are $x_1,x_2,x,\varphi$.
It is 
convenient to introduce the coordinates $y=x_1+ix_2,\bar y=x_1-ix_2$. 
As discussed in section 2.3, we have two possible
target space supersymmetries. If we choose
$(Q_{+,\alpha},\bar{Q}_{+,\alpha})$,$\alpha=1,2$, as the supercharges
we have the type IIB
superstrings with $(2,0)$ supersymmetry.
If we choose $(Q_{+,\alpha},\bar{Q}_{-,\alpha})$  as the supercharges
we get  type IIA superstrings with
$(1,1)$ supersymmetry.

Consider first the type IIB superstrings.
As before, we construct a bigger superspace.
We add the supercharges
 $(Q_{\dot +}^{\alpha},\bar{Q}_{\dot +}^{\alpha})$
defined in \supdot.

The superspace coordinates and their conjugate momenta are 
\eqn\gsA{
\t_{+}^{\a} = c \xi e^{- {3\over 2}\phi - {i \over 2} H^{1} \pm
{i\over 2} (H  + \sqrt{3} x)}\,, ~~~~~
p_{+, \a} = b \eta e^{ {3\over 2}\phi + {i \over 2} H^{1} 
\mp {i\over 2} (H + \sqrt{3} x)}\,, 
}
$$
\t_{\dot +}^{\dot \a} = e^{ {1\over 2} \phi + {i\over 2} H^{1} 
\pm {i\over 2} (H -   \sqrt{3} x)}\,, ~~~~~
p_{\dot +, \dot \a} = e^{ - {1\over 2}\phi - {i\over 2} H^{1}  
\mp {i\over 2} (H -   \sqrt{3}) x}\,. 
$$

The variables $\t_{+}^{\a}$ and $\t_{\dot +}^{\dot \a}$ 
have regular OPE, and 
\eqn\susya{
q_{+,\a}(z)\theta_{+}^{\beta}(w)  \sim  {\delta_{\a}^{\beta} 
\over (z-w)},~~~~~~
q_{\dot +,\dot \a}(z)\theta_{\dot +}^{\dot \beta}(w)  \sim  {\delta_{\dot{\a}}
^{\dot{\beta}} \over (z-w)}\,.
}
Also,
 \eqn\ptheq{
p_{+,\a}(z)\theta_{+}^{\beta}(w)  \sim 
 {\delta_{\a}^{\beta} \over (z-w)},~~~~~~
p_{\dot +, \dot \a}(z)\theta_{\dot +}^{\dot \beta}(w)  \sim  
{\delta_{\dot{\a}}^{\dot{\beta}} \over (z-w)}\,.
}

We define 
\eqn\xp{
x' = x + i \sqrt{3} (\phi + \kappa) \,
}
and the current
\eqn\Jc{
J_{x'} = \p x' = \p x + i \sqrt{3} (\p \phi + \p \kappa) \,. 
}
The current $J_{x'}$ commutes with \gsA. 
We have
\eqn\ope{
 J'(z) J'(w) \sim {2 \over (z-w)^{2}}\,. 
}
There is one chiral boson $\rho$ defined by 
\eqn\alF{
\rho = 3 \phi - \chi + 2\kappa - i \sqrt{3}  x\,, ~~~~~
}
and we have
\eqn\bg{
b =  {1\over 2} p_{+,\a}p_{+,\b} e^{\a\b} \e^{-\rho}, ~~~~~
\gamma^{2} b = {1\over 2}p_{\dot +,\dot \a}p_{\dot +,\dot \b} 
\e^{\dot \a \dot \b} e^{\rho}\,.
}
As before we define the current $J'$  
\eqn\Jg{
J' = - \p\rho - i \sqrt{3} J_{x'}\,,
}
which is the ghost current of the BRST symmetry.
\eqn\Je{
J' = - \p (  3 \phi - \chi + 2\kappa + i \sqrt{3} x) - i \sqrt{3} J_{x} = \p \kappa + \p \chi\
= cb + \eta \xi\,.
}

The stress tensor takes the form
\eqn\alG{
T'= - p_{+\a} \p \t^{+\a} - p_{\dot + \dot \a} \p \t^{\dot \a} 
- \half \p y \p \bar{y}  
}
$$   
~~~~~~~~~
- \half (\p\rho)^{2} - \half \p^{2} \rho
- \half (\p x')^{2} - i {\sqrt{3}\over 2} \p^{2} x'
- {1\over 2} (\p\phi)^{2} + {\sqrt{3}  \over 2} \p^{2} \phi \,.
$$
The contribution from the central charge are given by
$$
(-4)_{p_{+},\t^{+}} + (-4)_{p_{\dot +}, \t^{\dot +}} + (+1)_{y} + (+1)_{\bar{y}} + (1 - 9)_{x'} + (1 + 9)_{\varphi} + (1 + 3)_{\rho} =0\,.
$$
Next we define
\eqn\Gp{
G'^-= {1\over 2} p_{+,\a}p_{+,\b} \e^{\a\b} e^{-\rho}\,.
}
$T',J',G'$
are three
of the generators of the twisted $N=2$ superconformal algebra
\scA.
$G'^+$ is the BRST current.

Denote
\eqn\defin{
x_{1\dot 1} =  (\varphi - i x'),~~~ 
x_{2 \dot 1} = \bar y,~~~ 
x_{1 \dot 2} = - y,~~~ x_{2\dot 2} = (\varphi + i x')\,.
}
We
introduce the supersymmetric invariant variables 
$d_{+\a}, d_{\dot + \dot \a}$ and $\Pi_{\a\dot \a}$:
\eqn\ppiA{
\Pi_{1\dot 1} = \p x_{1\dot 1} 
- i (\t^{+ 1} \p \t^{\dot + \dot 1} + \t^{\dot + \dot 1} \p\t^{+ 1})\,, ~~~
\Pi_{1\dot 2} = \p \bar{y}
- i (\t^{+ 1} \p \t^{\dot + \dot 2} + \t^{\dot + \dot 2} \p\t^{+ 1})\,,}
$$
\Pi_{2\dot 1} = -\p y
- i (\t^{+ 2} \p \t^{\dot + \dot 1} + \t^{\dot + \dot 1} \p\t^{+ 2})\,, ~~~
\Pi_{2\dot 2} = \p x_{2\dot 2}
- i (\t^{+ 2} \p \t^{\dot + \dot 2} + \t^{\dot + \dot 2} \p\t^{+ 2})\,, 
$$

\eqn\pdA{
d_{+\a} = p_{+,\a} + i \t^{\dot + \dot \a} \p x_{\a\dot \a} - \half 
(\t^{\dot +})^{2}  \e_{\a\b}\p\t^{+ \b} + {1\over 4} \e_{\a\b} \t^{+ \b} \p (\t^{\dot +})^{2} \,,
}
$$
d_{\dot +\dot \a} = 
p_{\dot +,\dot \a} + i \t^{+ \a} \p x_{\a\dot \a} - \half 
(\t^{+})^{2} \e_{\dot \a\dot\b}\p\t^{\dot + \dot \b} + {1\over 4} \e_{\dot\a\dot \b} 
\t^{\dot + \dot \b} \p (\t^{+})^{2} \,,
$$
in terms of which the energy-momentum tensor can be 
written as follows
\eqn\ppB{
T' = - d_{+\a} \p \t^{+\a} - d_{\dot + \dot \a} \p \t^{\dot + \dot \a} - \half \Pi_{\a\dot \a} \Pi^{\a\dot \a}
- \half (\p\rho)^{2} 
- \half \p^{2} \left(\rho + \sqrt{3}\Pi_{1 \dot 1} \right)\,.
}
It can be checked that by using the above definitions 
the energy-momentum tensor can be rewritten in terms of free variables. 
The construction of the supersymmetry currents can be derived easily 
in a similar way.  

The OPE of $d$'s are given by 
\eqn\ppBA{
d_{+1}(z) d_{\dot +\dot 1}(w) \sim {\Pi_{1 \dot 1}\over (z-w) }\,, ~~~
d_{+1}(z) d_{\dot +\dot 2}(w) \sim {\Pi_{1 \dot 2}\over (z-w) }\,, }
$$
d_{+2}(z) d_{\dot +\dot 1}(w)  \sim {\Pi_{2 \dot 1}\over (z-w) }\,, ~~~
d_{+2}(z) d_{\dot +\dot 2}(w) \sim {\Pi_{2 \dot 2}\over (z-w) }\,, ~~~
$$
and analogously one can construct the supersymmetry generators. 
The total supersymmetry for 
open superstring is $(1,0)$ and in the case of closed 
superstring is $(1,1)$ or $(2,0)$ for the type IIA or type IIB. 
This also implies that the spectrum at the 
massless level is generated for the open string by a two component 
fermion, and for closed string (in the RR sector) by bilinears of those 
supersymmetry charges. 

Combining left and right sectors, we have for constant RR fields 
of IIB/A string theories the vertex operators 
\eqn\tBd{
V^{A}_{RR} = q_{\dot +,\dot \a} \bar q_{\dot -,\dot \a}\,, ~~~~~
V^{B}_{RR} = q_{\dot +,\dot \a} \bar q_{\dot +,\dot \a}\,, 
} 
where $\bar q_{\dot +,\dot \a}$ and $\bar q_{\dot -,\dot \a}$ 
are the right moving 
charges. The
coupling of the RR vertex operators
to the space-time RR fields strength $F^{\alpha\beta}$ is
\eqn\RRcd{
F^{{\dot + \dot -\dot \a \dot \a}}q_{\dot +,\dot \a}
 \bar q_{\dot -,\dot \a}\,, ~~~~~
F^{{\dot + \dot + \dot\a \dot \a}}q_{\dot +,\dot \a} 
\bar q_{\dot +,\dot \a}\,.
}
Thus we find four
RR degrees of freedom  both for IIA and IIB.
In Type IIB this corresponds to a 1-form (or its Hodge dual 3-form)
field strength in four dimensions.
In Type IIA this corresponds to a 0-form  (or its Hodge dual 4-form)
field strength and a self-dual 
2-form field strength
in four dimensions.

Let us construct now a twisted  $N=2$ algebra with $\hat{c}=-1$ which will be compared
with a similar algebra of the uncompactified sector
of string theory compactified on $CY_{3}$. 
We consider the $U(1)$ charge 
\eqn\ppC{
J' = -  \p \left(\rho + \sqrt{3}\Pi_{1 \dot 1} \right)\,, ~~~~
J'(z) J'(w) \sim {- 1 \over (z-w)^{2}}\,, ~~~~
T'(z) J'(w) \sim {1 \over (z-w)^{3}}\,.
}
Next we perform 
the similarity transformation generated by 
\eqn\Rt{
R = \sqrt{3} \oint \Pi_{1\dot 1} \p \rho\,,
} 
which removes the term $\half \sqrt{3} \Pi_{1\dot 1}$ in $T'$ and $J'$, but
does not change 
the central term of $J'$ and the central charge. Notice that the operator 
$R$ will modify also the superderivatives $d_{+\a}, 
d_{\dot + \dot \a}$ and the translation generators. However, it will preserve
 the 
commutation relation. We denote by an additional hat the new operators 
$T', J', d, \dots, \rightarrow \hat T, \hat J, \hat d, \dots$. Now, the form of the 
energy-momentum tensor and of $\hat J$ is exactly that of the
uncompactified part of $CY_{3}$.  
In terms 
on these new variables we can 
construct the N=2 superconformal algebra by 
adding the new generators
\eqn\GG{
\hat G^{-} = \hat d_{+\a} \hat d_{+\b} \e^{\a\b} e^{- \rho}\,,~~~~~~~~
\hat G^{+} = \hat d_{\dot +\a} \hat d_{\dot +\b} \e^{\a\b} e^{\rho}\,. 
}
To check the algebra only the commutation relation and 
the OPE's for the chiral boson $\rho$ are needed. So, the last 
step is to express the algebra in the original variables by 
performing the $R$ similarity transformation back 
$${\cal O}' = e^{R} \hat {\cal O} e^{-R}$$
where $\hat {\cal O} = (\hat T, \hat J,\hat G^{+},\hat G^{-})$. The 
form of the generators $G^{\pm'}$ is established by computing 
the complete expansion of the similarity transformation. Notice that 
due to presence of the exponential $e^{\pm \rho}$ in the 
definition of $\hat G^{\pm}$, the new $G^{\pm \prime}$ is definitely more 
complicated. 


\subsec{$D=5$, $n=2$} 

We have the bosons $(x^1,...,x^4, x,\varphi)$
 and 
the ghosts $\beta, \gamma$.
There are eight  fermions obtained by the 
RNS fermions $\psi^{i}$, the super-Liouville partner $\psi_{l}$ and 
the ghosts $b,c$.
We define four spinors $\t^{\a}_{+}, \t^{\a'}_{+}$ 
and $\t^{\dot\a}_{\dot +},\t^{\dot\a'}_{\dot +}$ and their conjugates 
\eqn\odE{
\t^{a}_{+} = c \xi e^{-{3\over 2} \phi 
\pm {i\over 2}(H^{1} +H^{2}) 
+ {i \over 2} (H + \sqrt{2} x)}\,,
~~~~~~
p_{+,a} = b\eta \e^{{3\over 2} \phi \mp {i\over 2}(H^{1} +H^{2}) 
- {i \over 2} (H + \sqrt{2} x)}\,,
}
$$
\t^{\a'}_{+} = c \xi e^{-{3\over 2} \phi 
\pm {i\over 2}(H^{1} -H^{2}) 
- {i \over 2} (H + \sqrt{2} x)}\,,
~~~~~~
p_{+,\a'} = b\eta \e^{{3\over 2} \phi \mp {i\over 2}(H^{1} - H^{2}) 
+ {i \over 2} (H + \sqrt{2} x)}\,,
$$
$$
\t^{\dot\a}_{\dot +} = 
e^{{1\over 2} \phi 
\mp {i\over 2}(H^{1} + H^{2}) 
+ {i \over 2} (H - \sqrt{2} x)}\,, 
~~~~~~
p_{\dot +,\dot \a} = e^{{1\over 2} \phi 
\pm {i\over 2}(H^{1} + H^{2}) 
- {i \over 2} (H - \sqrt{2} x)}\,,
$$
$$
\t^{\dot\a'}_{\dot +} = 
e^{{1\over 2} \phi 
\mp {i\over 2}(H^{1} - H^{2}) 
- {i \over 2} (H - \sqrt{2} x)}\,, 
~~~~~~
p_{\dot +,\dot \a'} = e^{{1\over 2} \phi  
\pm {i\over 2}(H^{1} - H^{2}) 
+ {i \over 2} (H - \sqrt{2} x)}\,.
$$
However, they are not independent.
This is the same situation as 
in \BerkovitsIM. In order to display the full supersymmetry, one 
needs to add some auxiliary variables and to impose some new constraints 
\BerkovitsDU. 

By the counting of bosonic variables, we find that we need  chiral bosons
$\rho$ and $\chi$, where
\eqn\chiral{
\rho = 2 \phi + \kappa - i \sqrt{2} x\,,
}
and the chiral boson $\chi$ coincides with the original chiral boson of 
the bosonization of $b,c$-system. 
We define
\eqn\x{
x'= x + i \sqrt{2} (\phi + \kappa)\,,
}
and the current $J_{x'}$  
\eqn\cicA{
J_{x'} = \p x + i \sqrt{2} (\p \phi + \p \kappa)\,.
}
They combine into 
\eqn\current{
J' = - \p \rho + \p \chi - i \sqrt{2} J_{x} = \p \chi + \p \kappa\,,
} which is 
the ghost current of the $N=2$ superconformal algebra related to the BRST symmetry. 
With these definitions, it is easy to check that 
\eqn\cicB{
{1\over 4!} p_{+,a}p_{+,b}p_{+,c}p_{+,d} \e^{abcd} e^{2 \rho - \chi} = \gamma^{2} b\,.
}

The total conformal charge is obtained by summing the contributions 
coming from the bosonic variables $x^{i}$ and $\vp$, which give 
$5 + 7$, the fermionic variables which yield $-8$, and the chiral bosons 
whose contribution which is $-4$. Again the contribution of the chiral bosons 
to the twisted conformal charge matches the contribution of the 
chiral boson for the compactification of critical superstring compactified
on $CY_{2}$. 

 By twisting 
the theory with the currents $J = - \p \rho$, we have that the model 
can be compared  with the critical superstring on $CY_{2}$ and 
therefore similar analysis to the one performed in \BerkovitsIM\ can be repeated. In 
\BerkovitsIM\ only half of the supersymmetry is manifest and in a 
subsequent paper  a formalism \BerkovitsDU\ 
with manifest N=2 supersymmetry of the target space is constructed by 
duplicating the number of variables and imposing an harmonic constraint. 

Here, we have to perform a similar construction, in order to have 
the manifest supersymmetry. We  duplicate the number of 
variables (adding 8 fermionic variables $\t^{a \prime}$ and 
their conjugate momenta) and pick up only those which are local 
with respect to each others. In addition, one has to 
check that the harmonic constraint is compatible 
with this choice. 

This leads to 
a superspace, which implements the $N=1$ supersymmetry for D=4 (for 
the open superstring) and $N=(1,1)$ or $(2,0)$ for type IIA/B. Out 
of the 8 variables $(\t^{a}, \t^{a \prime})$, one selects 
four fermionic variables of the four dimensional superspace and 
the algebra of covariant derivatives and supercharges is 
the usual one. 

Note the following interesting aspect: 
we have mapped the original variables of RNS string theory 
into the bosonic variables 
$x^{i}, x', \varphi, \rho, \chi$ and into the fermionic  variables 
$\theta_{a}, p^{a}$. For a generic background these variables are 
entangled in the sigma model. However, if the background has
a factorized structure where 
some of the variables are not mixed, we can have a simplified situation: 
the variables $x^{i}, \t_{a}, p^{a}, \rho$ form a $N=2$ superconformal 
system by their own with $\hat c = -1$. The rest of the variables $x', 
\varphi, \chi$ form an $N=2$ superconformal algebra with 
$\hat c = 3$.\foot{We thank 
G. Policastro and T. Dasgupta for discussion on this point.} 
 Using this framework the analysis of such systems 
is simplified. 

\subsec{$D=7$, $n=3$}

This is  the last interesting example of this
class of non-critical superstrings. The 
transverse space is parametrized  by six coordinates $x^{[ab]}$ 
(with $a=1,\dots,4$) and the longitudinal space is generated by 
$x', \varphi$, where $x' = x + i (\phi + \kappa)$. The original 
fermions together with bosonization of the ghost fields 
cannot be mapped into the 16 $\theta_{i}, p^{i}$ (with $i=1, \dots,8$). 
In order to have the manifest supersymmetry one has to enlarge the 
space by addition fermionic variables (such as in  \BerkovitsDU\ and
\BerkovitsVN). The present situation is even more 
complicated than the D=5 case and we suspect that 
an analysis using pure spinor might simplify the analysis \op. 


\newsec{Non-critical strings in curved target space}

In this section we consider non-critical
strings on curved target spaces with Ramond-Ramond background fields 
in the hybrid type formalism.

\subsec{Coupling to Ramond-Ramond fields}

A simple and local coupling of the worldsheet fields to the RR 
background is a common fundamental feature of the hybrid formalism and of the
pure spinor formalism. Let us briefly review some basic facts and  
explain the structure of the couplings and the RR vertex operators. 
In the hybrid formalism, we add to
the superspace coordinates $(x^{m}, \t^{\a})$ the conjugate momenta
$p_{\a}$. 
In order to respect the target space 
supersymmetry we form the supersymmetric invariant quantities 
$\Pi^{m}, \bar{\p} \t^{\a}, d_{\a}$ and
$\bar{\Pi}^{m}, \p \bar{\t}^{\a}, \bar{d}_{\a}$, which
are worldsheet  holomorphic or anti-holomorphic
1-forms. 
They commute with the supersymmetry charges
$Q_{\a} = \oint q_{\a}$, where $q_{\a}$ are the supersymmetric currents.  
   
At the massless level, the vertex operator is constructed in terms of these 
fundamental blocks
\eqn\RRvA{
V = 
\p \t^{\a} \p \bar\t^{\beta} A_{\a\b} + 
\Pi^{m} \p \bar\t^{\beta} B_{m \b} + 
\p \t^{\a} \bar\Pi^{n} B_{\a n} +
\Pi^{m} \bar\Pi^{n} C_{mn} +
}
$$
d_{\a}  \p \bar\t^{\beta} D^{\a}_{~\b} + 
\p \t^{\a} \bar d_{\b} D_{\a}^{~\b} +
d_{\a}\bar \Pi^{n} E^{\a}_{~n} +
\Pi^{m} \bar d_{\b} E_{n}^{~\b} +
d_{\a} \bar d_{\b} F^{\a \b} + \dots \,.
$$
where $A_{\a\b}, \dots, F^{\a\b}$ are superfields. 
The ellipsis stand for the additional contributions 
coming from the ghost fields. The form of these couplings 
relies on the dimension of the spacetime, the $R$-symmetry 
and upon the Lorentz transformation properties of the ghost fields. 

The lowest components of the superfield $C_{mn}$ are the graviton, 
the NS-NS 2-form and the dilaton
\eqn\dilaton{
C_{mn} = g_{mn} + b_{mn} + \eta_{mn}\phi + O(\t,\bar \t)\,.
}
The lowest component of the superfield  $F^{\a \b}$
is the RR field strength (in spinor indices)
\eqn\ramondramond{
F^{\a \b} =f^{\a \b} + O(\t,\bar \t)\,.
}
For a complete discussion see \tama. 
By imposing the equation $\{Q_B, V\} = \p U$ where $Q_B$ is the 
BRST charge and 
$U$ is generic vertex operator with ghost number $(1,0)$ and 
conformal spin $(0,1)$ (as is explained in \tama) one gets superspace 
relation among the different superfields.
In the case of constant RR field strengths
(which are solutions of the linearized supergravity equations), 
\eqn\cosntramon{
F^{\a\b}(x,\t, \tilde \t) = f^{\a\b}\,,
}
and 
the expression \RRvA\ reduces to 
\eqn\RRvB{
V = f^{\a\b} q_{\a} \bar q_{\b}\,. 
}
This gives a local coupling between the RR field strengths and 
the supersymmetry currents. For constant RR fields all contributions 
of the superfields to the vertex operators conspire to give the supercurrents 
even if the coupling has been written starting with the $d$'s. 
Note that in the same way one obtains the 
coupling to Ramond fields in the case of open superstrings. 

The 
form of the 
supersymmetry currents is usually non-linear in the 
worldsheet fields and, generically, by adding the vertex operators \RRvB\ 
it is not easy to compute the contribution of the RR field to the amplitudes 
exactly. However, all propagators of the sigma model are well-defined and 
in special cases (constant RR fields) the fields $p_{\a}$ can be integrated 
out easily. If we wish to study the sigma models with the addition 
of these deformations, we have to add also the back-reaction as will be discussed
later.

The $(2n+2)$-dimensional
target space even forms RR field strengths of type IIA are encodes
in $F_{\a\b}$ via
\eqn\RRfs{
F_{\a}^{~\b} = \delta_{\a}^{~\b}F^{(0)} + {1 \over 2!} (\gamma^{mn})_{\a}^{~\b}F_{mn}
+{1 \over 4!} (\gamma^{mnpq})_{\a}^{~\b}F_{mnpq}\,,
}
and  the odd forms RR field strengths of type IIB as
\eqn\RRfsb{
F_{\a\b} =  \gamma_{\a\b}^{m}F_{m}
+ {1 \over 3!} \gamma_{\a\b}^{mnp}F_{mnp}
+{1 \over 5!} \gamma_{\a\b}^{mnpqr}F_{mnpqr}\,.
}
The gamma matrices used in the above equations are the off-diagonal 
$16 \times 16$ blocks of ten-dimensional Dirac matrices 
$(\Gamma^{0} \Gamma^{m})_{\a\b}$. 
They are real and symmetric and they satisfy the Fierz identities 
$\gamma^{m}_{(\a\b} \gamma_{m\g\d)} =0$. 
Note that the forms that appear are those whose degree is not higher than the
target space dimension. By dimensional reduction one finds all lowest 
dimensional models with their RR fields couplings.

\subsec{Target space effective action}

As we have seen, the RR field strengths of non-critical type IIB and type IIA superstrings
are odd forms and even forms respectively.
In addition, in $d=2n+2$ dimensions the middle $(n+1)$-form is self-dual.
This is quiet different than the structure of RR fields 
of the critical type II superstrings compactified on a $CY_{4-n}$ manifold. 

Let us review the counting of the RR degrees of freedom.
In $d=2$ there is one RR degree of freedom.
In type IIB it is 
a self-dual 1-form and in type IIA
a scalar (or its Hodge dual 2-form).
In $d=4$ there are 4  RR degrees of freedom.
In type IIB it a 1-form (or its Hodge dual 3-form) and
in type IIA it is 
a self-dual 2-form.
In $d=6$ the there are 16  RR degrees of freedom.
In type IIB these are a 1-form (or its Hodge dual 5-form) and
a self-dual 3-form, and  in type IIA 
 these are 0-form  (or its hodge dual 6-form) and a 2-form
 (or its hodge dual 4-form).
In $d=8$ the there are 64  RR degrees of freedom.
In type IIB these are a 1-form (or its Hodge dual 7-form) and
a 3-form  (or its Hodge dual 5-form), and  in type IIA 
 these are 0-form  (or its Hodge dual 8-form) and a 2-form
 (or its Hodge dual 6-form) and a self-dual 4-form.

As discussed before, unlike the 
critical superstrings case, 
the low energy approximation $E\ll l_s^{-1}$ is not valid.
The reason being that the action
contains a cosmological constant type term 
which vanishes only for $d=10$,
and the higher order curvature terms $\left(l_s^2{\cal R}\right)^n$ 
cannot be discarded.
One can still write an action for the massless fields,
whose bosonic part takes the form
\eqn\effec{
S = {1 \over 2 k_d^2} \int d^d x \sqrt{G}\left(e^{-2 \Phi}\left(R + 4(\p \Phi)^2 + 
{10 -d \over \alpha'} - {1 \over 2\cdot 3!}H^2\right) - {1 \over 2 \cdot n!} 
F_n^2 \right)\,.
}  
However, solutions to the field equations
have string scale curvature. 
For instance, consider curved backgrounds with RR fields, with constant dilaton and
vanishing 
NS-NS field, which will
be considered later. 
Then, the field equations of \effec\ imply that the scalar curvature
is
\eqn\curvature{
l_s^2{\cal R} = d-10\,.
}
One class of such backgrounds of type IIA non-critical
strings are $AdS_d$ spaces with a constant dilaton
$e^{2\Phi} = {1 \over N_c^2}$ and a $d$-form RR field $F_d$ 
\eqn\rr{
l_s^2F_d^2 = 2(10-d)d! N_c^2\,.
}

We note that even though consistent backgrounds of non-critical strings
may be solutions of the field equations of the action \effec,
the analysis of fluctuations is likely to give a wrong 
spectrum.

\subsec{D=1, n=0}

Using the supersymmetric variables the classical action for IIB in the flat background is given 
by 
\eqn\actA{
S_{IIB} = {1\over \a'} \int dz d\bar z \Big( 
\half \Pi_{+\dot +} \overline\Pi^{+\dot +}  + 
d_{+} \bar\p\t^{+} + d_{\dot +} \bar\p\t^{\dot +} + 
\bar d_{+}  \p \bar \t^{+} + \bar d_{\dot +} \p \bar\t^{\dot +} 
\Big) + S^{flat}_{B} \,,
}
where $S_{B}$ is the action for the chiral bosons $\omega,\bar{\omega}$ 
and $\rho,{\bar \rho}$.
We also introduced the right-moving sector. As explained above the 
choice of the right-hand supersymmetry charge determines if the 
closed model is IIA or IIB. 
The the classical action for IIA in the flat background takes the form 
\eqn\actAtd{
S_{IIA} = {1\over \a'} \int dz d\bar z \Big( 
\half \Pi_{+\dot +} \overline\Pi^{+\dot +}  + 
d_{+} \bar\p\t^{+} + d_{\dot +} \bar\p\t^{\dot +} + 
\bar d_{-}  \p \bar \t^{-} + \bar d_{\dot -} \p \bar\t^{\dot -} 
\Big) 
+ S^{flat}_{B} \,,
}
On flat Riemann surface, there is no coupling 
with the background charge. 

In order to couple the system to the background,
we introduce the curved vielbeins $E^{A}_{~M}$ where the $A$ are tangent superspace
indices and $M$ are curved superspace indices.
We will use the notation introduced in \xpp\
 $({}_{+\dot +},{}^{+\dot +}, +, \dot +)$ for the tangent superspace indices,
and $Z^M$ for the curved target superspace coordinates.
The
the new supersymmetric 
variables are given by 
\eqn\variables{
\Pi^{A} = E^{A}_{M} \p Z^{M},~~~\overline \Pi^{A} = E^{A}_{M} \bar\p Z^{M}\,.
}
In terms of the vielbeins, on can derive the 
the covariant derivatives $D^{A} =  (E^{-1})^{A}_{~M} D^{M}$ where 
$E^{-1}$ is the inverse of the vielbeins and $D^{M}$ are the 
covariant derivatives established in section 3.1. 
We also introduce the 
NS-NS 2-form $B_{AB}$ using the superfield
\eqn\actAB{
(G +B)_{AB} = E^{M}_{~A} E_{B M}\,.
}
The R-R fields are
$F^{\dot + \dot +}$ for type IIB and $F^{\dot + \dot -}$ for type IIA strings.

The action for IIB in curved space can be written as
\eqn\actAC{
S_{IIB} = {1\over \a'} \int dz d\bar z \Big( 
(G+B)_{AB} \Pi^{A} \overline\Pi^{B}\ + 
d_{+} \bar{\Pi}^{+} + d_{\dot +} \bar{\Pi}^{\dot +} +
\bar d_{+} \Pi^{+} + \bar d_{\dot +}  \Pi^{\dot +}  }
$$
+ d_{\dot +} \bar d_{\dot +} F^{\dot + \dot +}\Big)
+ S_{B}\,.
$$
Similarly, the action for type IIA takes the form
\eqn\actAC{
S_{IIA} = {1\over \a'} \int dz d\bar z \Big( 
(G+B)_{AB} \Pi^{A} \overline\Pi^{B}\ + 
d_{+}  \bar{\Pi}^{+} + d_{\dot +}  \bar{\Pi}^{\dot +} +
\bar d_{-} \Pi^{-} + \bar d_{\dot -}  \Pi^{\dot -}  }
$$
+ d_{\dot +} \bar d_{\dot -} F^{\dot + \dot -}\Big)
+ S_{B}\,.
$$

$S_{B}$ is the action for the chiral bosons $\rho$ and $\omega$. 
Note that there is no worldsheet-covariant formulation 
for chiral bosons and the action should be supplemented by the chirality 
condition. 
In order to write
the  action $S_{B}$ for the chiral bosons 
we notice that
the field $\rho$ depends on $x$ (see equation \cbC) and therefore  
 couples to the $U(1)$ connection $A^{R}_{M}$ 
 of the R-symmetry \J. The
action $S_{B}$ reads 
 \eqn\actAD{
S_{B} = S_B^{flat} +\int dz d\bar z
\left( \bar\p Z^{M}A^{R}_M \p\rho + 
\p Z^{M}
A^{R}_M \bar{\p}\bar\rho \right)\,.
}
So far we have been
discussing the the classical sigma model. However, 
in order for the model to be conformally invariant, 
one needs to add  a Fradkin-Tseytlin term. Following 
\BerkovitsTG\ 
we see that the additional term we should add is 
of the form 
\eqn\FT{
S_{FT} = \int d^{2}z D_{+} \bar D_{+} (\hat \Phi_{cc} \Sigma_{cc} + h.c.)\,,
}
where $\hat \Phi_{cc}$ is a $N=2$ chiral superfield whose lowest component 
contains the spacetime dilaton. 
The superfield $\Sigma_{cc}$ is a chiral 
superfield and the ordinary worldsheet curvature is given 
by $D_{+}D_{-} \Sigma_{cc}$ where $D_{\pm}$ are the N=2 worldsheet 
superderivatives. 
When the dilaton is constant we get from
\FT\ the Euler number
 of the Riemann 
surface. 
\subsec{$AdS_2$}

Consider the example  of $AdS_{2}$ background of type IIA 
non-critical string studied in
\Verl\ in the Green-Schwarz formalism.
Let $Z,\bar Z$ denote the coordinates on  $AdS_{2}$.
The dilaton $\Phi$, the metric $G$ and the RR 2-form $F$ take the form
\eqn\background{
e^{2\Phi} = {1 \over N_c^2},~~~G_{Z\bar Z} = -{1 \over 2(Z-\bar Z)^2},~~~
F_{Z\bar Z} = {8N_c \over (Z-\bar {Z})^2}\,.
}

We denote the curved superspace
coordinates by $Z_{M} = (Z,\bar Z, \Theta^{\dot +}, \bar\Theta^{\dot -})$.
In addition there are two free variables 
$(\Theta^{+}, \bar\Theta^{-})$ needed for the extension
of the superspace.  
The tangent space
coordinates are denoted by $z_{A} = (z, \bar{z}, \theta^{\dot +}, \bar{\theta}
^{\dot -})$, and in addition we have
$(\theta^{+}, \bar\theta^{-})$.
For the simplicity of the notation
we denote  $\Theta = \Theta^{\dot +}, \bar\Theta =\bar\Theta^{\dot -}$ and
 $\theta = \theta^{\dot +}, \bar\theta =\bar\theta^{\dot -}$.
Note that we use the same symbols $(z, \bar{z})$
for the target space tangent coordinates and the
worldsheet coordinates.  
The curved quantities $\Pi^{A}$ are related to the flat ones by the 
vielbeins $\Pi^{A} = E^{A}_{M} \p Z^{M}$ where 
\eqn\viel{
E^{z}_{~Z} = {1\over (Z- \bar Z - \Theta \bar\Theta)}\,, ~~~~~~
E^{z}_{~\Theta} = 
{\Theta \over (Z- \bar{Z} - \Theta \bar\Theta)}\,, ~~~~~~
}
$$
E^{\theta}_{~Z} = {\Theta -\bar\Theta
\over  (Z- \bar Z - \Theta \bar\Theta)^{3/2} }\,, ~~~~~~
E^{\theta}_{~\Theta} = {1\over (Z - \bar Z - \Theta \bar \Theta)^{1/2}} 
- {\Theta \bar\Theta 
\over  (Z- \bar Z - \Theta \bar\Theta)^{3/2}}\,. ~~~~~~
$$
Together with these vielbeins there are also the conjugated 
$
E^{\bar z}_{~\bar Z}, 
E^{\bar z}_{~\bar\Theta}, 
E^{\bar \theta}_{~\bar Z}, 
E^{\bar \theta}_{~\bar\Theta}$ we are deduced in the same 
way as \viel\ from the Maurer-Cartan forms of the $Osp(1,2)$ algebra. 

The action takes the form
\eqn\gd{
S_{IIA} = \int d^{2}z  \Big(\Pi_{z} \bar \Pi_{\bar z}+
d_{\dot +} E^{\t}_{M} \bar\p Z^{M} +  
d_{+} \bar\p \t^{+} +
\bar d_{\dot -} E^{\bar\t}_{M} 
 \p \bar Z^{M}  + \bar d_{-} \p \bar\t^{-}  + F^{\dot + \dot -}
d_{\dot +} \bar{d}_{\dot -}
\Big)
}
where $F^{\dot + \dot -} = 8 N_{c}$ is the constant 
RR field strength. 

In addition we have the chiral boson action $S_B$.
In order to write it we need the
$U(1)_R$ connection, which in this case is replaced
by the dilatation (there is no $U(1)_R$ part in the $OSp(1,2)$ algebra).
It takes the form
\eqn\rcharge{
A ={d Z + d \bar{Z} + \bar{\Theta} d \Theta + \Theta d \bar{\Theta}
\over Z-\bar Z - \Theta \bar \Theta}\,.
}
Since the dilaton is constant we get from
\FT\ the Euler number of the Riemann 
surface. 
Note also that  $d_{+}, d_{\dot +},\bar d_{-}, \bar d_{\dot -}$ 
can be integrated
out easily. 
We will leave the complete analysis for a 
forthcoming publication.  

\subsec{D=3, n=1}

Using the supersymmetric variables the classical action for IIB in the flat background is given 
by 
\eqn\actBfd{
S_{IIB} = {1\over \a'} \int dz d\bar z \Big( 
\half \Pi_{\a\dot \a} \overline\Pi^{\a\dot \a}  + 
d_{+\a} \bar\p\t^{+\a} + d_{\dot +\dot\a} \bar\p\t^{\dot +\dot\a} + 
\bar d_{+\a}  \p \bar \t^{+\a} + 
\bar d_{\dot +\dot\a} \p \bar\t^{\dot +\dot\a} 
\Big) }
$$
+ S^{flat}_{B} \,,
 $$
where $S_{B}$ is the action for the chiral bosons
$\rho,{\bar \rho}$.

The the classical action for IIA in the flat background takes the form 
\eqn\actAfd{
S_{IIA} = {1\over \a'} \int dz d\bar z \Big( 
\half \Pi_{\a\dot \a} \overline\Pi^{\a\dot \a}  + 
d_{+\a} \bar\p\t^{+\a} + d_{\dot +\dot\a} \bar\p\t^{\dot +\dot\a} + 
\bar d_{-\a}  \p \bar \t^{-\a} + \bar d_{\dot -\dot\a} \p \bar\t^{\dot -\dot\a} 
\Big) }
$$
+ S^{flat}_{B} \,.
 $$
As before, on flat Riemann surface, there is no coupling 
with the background charge.

In order to couple the system to the background,
we introduce the curved vielbeins $E^{A}_{~M}$ where the $A$ are tangent superspace
indices and $M$ are curved superspace indices.
We will use the notation $x_{\a \dot \a}$ \defin.

The action for IIB in curved space can be written as
\eqn\actACfd{
S_{IIB} = {1\over \a'} \int dz d\bar z \Big( 
(G+B)_{AB}  \Pi^{A} \overline\Pi^{B}\ + 
d_{+\a} \bar \Pi^{+\a} + d_{\dot +\dot\a} \bar \Pi^{\dot +\a} +
\bar d_{+\a} \Pi^{+\a} + \bar d_{\dot +\dot \a} \Pi^{\dot +\dot\a} }
$$
+ d_{\dot +\dot \a} \bar d_{\dot +\dot \a} F^{\dot + \dot + \dot \a \dot \a}
 \Big)+ S_{B}\,.
$$
Similarly, the action for type IIA takes the form
\eqn\actACfd{
S_{IIA} = {1\over \a'} \int dz d\bar z \Big( 
(G+B)_{AB} \Pi^{A} \overline\Pi^{B}\ + 
d_{+\a} \Pi^{+\a} + d_{\dot +\dot \a}  \Pi^{\dot + \dot \a} +
\bar d_{-\a} \bar \Pi^{-\a} +
 \bar d_{\dot -\dot \a} \bar \Pi^{\dot -\dot \a}  }
$$
+ d_{\dot + \dot \a} \bar d_{\dot - \dot \a} F^{\dot + \dot - \dot \a \dot \a}
\Big) + S_{B}\,.
$$
Again we have to establish the form of the action $S_{B}$ in both 
cases. In a separate publication \op, we will explore in 
detail the structure of chiral primary fields and 
the constraints on the sigma model due to the $N=2$ superconformal symmetry.

The invariance 
of the action under the superconformal transformations implies the form of the 
chiral boson couplings. In addition, we should add the FT term $S_{FT}$ 
in order to guarantee the conformal invariance also at higher 
orders in $\a'$. As in the hybrid formalism \BerkovitsCB\ 
the form of the FT term is 
\eqn\FTA{
S_{FT} = \int d^{2}z D_{+} \bar D_{+} (\hat \Phi_{cc} \Sigma_{cc} + h.c.)\,,
}
where $\hat\Phi$ is the conformal compensator whose 
lowest component contains the dilaton field. 

\subsec{Higher dimensions}
 
 In the previous section 
we established the relation between the original variables and the 
hybrid formalism variables.
In order to have  the manifest supersymmetry we
use the left-movers 
$\Theta^{+ a}, \Theta^{\dot + \dot a}$ (with $a,a'=1, \dots,4$), 
the right-movers 
$\bar{\Theta}^{+ a}, {\bar\Theta}^{\dot + \dot a}$ and their conjugate 
momenta $\Delta$'s.
The type IIA $\sigma$-model takes the form 
\eqn\actACfdhigher{
S_{IIA} = {1\over \a'} \int dz d\bar z \Big( 
(G+B)_{AB} \Pi^{A} \overline\Pi^{B}\ + 
d_{+ a} \bar \Pi^{+ a} + d_{\dot +\dot a}  \bar \Pi^{\dot + \dot a} +
\bar d_{- a}  \Pi^{- a} + \bar d_{\dot -\dot a}  \Pi^{\dot -\dot a}  
}
$$
+ d_{\dot + \dot a} \bar d_{\dot - \dot b} F^{\dot + \dot - \dot a \dot b} +
\Delta_{+ a} \bar \p\Theta^{+ a} + 
\Delta_{\dot + \dot a} \bar \p\Theta^{\dot + \dot a} +
\bar\Delta_{+ a} \p\bar\Theta^{+ a} + 
\bar \Delta_{\dot + \dot a} \p \bar \Theta^{\dot + \dot a} \Big) + 
S_{B} + S_{FT}\,.
$$
Again, the chiral boson action $S_{B}$ and the FT term $S_{FT}$ \FTA\
are established using the $N=2$ superconformal 
invariance.
The chiral bosons couple to the 
worldsheet R-symmetry and to the Lorentz connection accordingly. 

In addition, we should also impose an harmonic constraint
to remove the doubling variables $\Theta$'s and $\Delta$'s. The form 
of these harmonic constraint is discussed in \BerkovitsDU.

Note that some simplifications arise when studying the class of curved 
backgrounds with a constant dilaton and RR flux such as, for instance,
$AdS_{2p}$ or  $AdS_3\times S^3$. 
Since the dilaton is constant the 
 FT term $S_{FT}$ is simply the Euler number of the Riemann surface.
The RR field couples as a constant in the action
\actACfdhigher\ and $d's$ and $\bar d's$ can be integrated out easily.
The rest of the action
is determined by the supergroup structure.

\subsec{The inclusion of open strings}

In order to enlarge the possible conformal backgrounds of non-critical
strings, open strings can be included. In
\KlebanovYA\ a Born-Infeld type term corresponding to
$N_f$ branes-antibranes uncharged system has been added
\eqn\binfeld{
S_{open} =  {-2N_f \over 2 k_d^2} \int d^d x \sqrt{G} e^{- \Phi}\,,
}
which allows for gravity solutions such as $AdS_5 \times S^1$.
In our framework, such a term is generated
by considering worldsheets with boundaries.

The inclusion of open string vertex operator can be done 
in the same way as for the closed deformations. The only difference 
is that the vertex operator has to placed on the boundary of the worldsheet. 
The general form of the massless boundary vertex operator is 
\eqn\openA{
V_{open} = \oint dz (\p\t^{\a} A_{\a} + \Pi^{m} A_{m} + d_{\a} W^{\a} + \dots)
}
 where $A_{\a}, A_{m}, W^{\a}$ are superfields. The ellipsis stands for 
 the ghost contributions and they depend upon the dimension 
 of the space time. The lowest component of the 
 superfields $A_{m}$ is represented by the gluon field and 
 $W^{\a} = \psi^{\a} + \dots$ has the gluino as the lowest component. 
 In the D=1 case, the only massless vertex operator 
 which is independent of Liouville field is a constant gluino field 
 and the coupling reduces to 
 \eqn\openB{
 V_{open} = \oint \psi^{\a} q_{\a}\,,
 }
where $q_{\a}$ is the supercurrent. 

Given the vertex operator for the massless sector, we can 
construct the sigma model in  curved space. One starts from 
IIB case (where the same supercharges are taken in the 
left- and right-moving sector) and impose the boundary conditions 
(for the flat case) at the level of superspace variables
\eqn\bouc{
(\t^{+} - \bar \t^{+})_{|_{z=\bar z}} = 0\,, ~~~~~~
(\t^{\dot +} - \bar \t^{\dot +})_{|_{z=\bar z}} = 0\,, 
}
$$
(p_{+} - \bar p_{+})|_{z=\bar z} = 0\,, ~~~~~~
(p_{\dot +} - \bar p_{\dot +})|_{z=\bar z} = 0\,, 
$$
(the usual boundary conditions are imposed on bosonic 
coordinates). At the level of sigma model one has to: 
{\it i}) construct a supersymmetric sigma model 
by adding the surface terms in order to compensate those 
supersymmetry variation which vanish because of partial integration, 
{\it ii}) 
add the vertex operator \openB\ and derive the Dirac-Born-Infeld action 
as a consequence of the boundary conditions in the given background. 


 \newsec{Discussion}
 
 In this work we constructed a hybrid type
formalism in order to study 
 non-critical strings in a manifest space-time
supersymmetric way.
We started with the linear dilaton background and worked
out the precise map of the RNS variables
to the supersymmetric variables.
We noticed that
 in order to construct manifest supersymmetric 
 non-critical string theory, we needed to double the superspace coordinates. 
This suggests that the non-critical strings may have
consistent backgrounds with double the number of supersymmetries
of the linear dilaton background.

One of the ingredients of the present formalism is the 
 presence of new ghost fields represented by a set of chiral 
 bosons.
 The relation between the original set of variables and the 
 new GS-like variables determines also the coupling of the ghost 
 fields to the supergravity background in the curved space. 

A feature of the present framework   is 
the possibility to couple the $sigma$-model to curved backgrounds 
and it
provides simple way to couple the RR fields to the worldsheet field.
This allows us, in particular, 
to study the conformal invariance to all orders in $\a'$.

We have
seen 
 several similarities of the construction
with that of Calabi-Yau compactifications of 
the critical superstrings, though the systems are different.

There are 
numerous open problems that should be addressed. Let us mention a few of 
them.
({\it i}) The analysis of the superconformal invariance 
of the sigma model for curved backgrounds.
({\it ii}) The 
analysis of the spectrum for higher dimensional examples extending 
the results of the lowest dimensional case.
({\it iii}) The construction 
of tree level scattering amplitudes and the higher genus extension.
({\it iv}) The relation with pure spinor formulation in ten dimensional 
superstring and its dimensional reduction.
({\it v}) The analysis of 
specific curved background such as AdS$_{2n}$, AdS$_{5} \times$S$^{1}$
(with open strings) and others.

\vskip 1cm
\noindent
\centerline{\bf Acknowledgments}
\medskip
We would like to thank I. Adam, O. Aharony, T. Dasgupta, A. Giveon,
D. Kutasov, M. Porrati, G. Policastro and J. Sonnenschein
for 
valuable  discussions.
P.A.G. is grateful to Tel-Aviv University where part of this work 
has been accomplished.  Y.O. would like to thank CERN, Theory Unit, where 
this work started. 


\newsec{Appendix}

\subsec{Supersymmetry currents and charges}

For the reader convenience, we add the relevant OPE's for the 
supersymmetry currents. For $D=1, n=0$ case, we have the currents 
\eqn\suchaA{
q_{+} = e^{- \half \phi -{i\over 2} (H + 2 x)}\,, ~~~~
q_{-} = e^{- \half \phi +{i\over 2} (H + 2 x)}\,, ~~~~
}
$$
q_{\dot +} = e^{- \half \phi - {i\over 2} (H - 2 x)}\,, ~~~~
q_{\dot  -} = e^{- \half \phi + {i\over 2} (H - 2 x)}\,. ~~~~
$$
which are nilpotent and satisfy
$$
q_{-}(z) q_{\dot -}(w) \sim {1\over z-w} e^{-\phi + i H}\,,~~~~~~
q_{-}(z) q_{\dot +}(w) \sim 0\,,
$$
\eqn\suchaB{
~~~~~~
q_{-}(z) q_{+}(w) \sim {1\over (z-w)^{3/2}} e^{-\phi}\,,
~~~~~~~~~~~~
q_{\dot +}(z) q_{\dot -}(w) \sim {1\over (z-w)^{3/2}} e^{-\phi}\,,
}
$$
q_{\dot +}(z) q_{+}(w) \sim {1\over z-w} e^{-\phi - i H}\,,
~~~~~~
q_{+}(z) q_{\dot -}(w) \sim 0\,. 
$$
As explained in the text, we can use only the set of 
supercharges $(\oint q_{+}, \oint q_{\dot +})$ or $(\oint q_{-}, \oint q_{\dot -})$ 
to construct the supersymmetric model. 

In the D=3, n=1 case we have eight combinations. 
\eqn\sugraC{
q^{\a}_{+} = e^{-\half \phi + {i\over 2} H^{1} \mp {i\over 2} (H + \sqrt{3} x)}\,, 
~~~~~~
q^{\a}_{-} = e^{-\half \phi - {i\over 2}
 H^{1} \pm {i\over 2} (H + \sqrt{3} x)}\,, 
}
$$
q^{\dot \a}_{\dot +} = e^{-\half \phi -  {i\over 2}
H^{1} \pm {i\over 2} (H - \sqrt{3} x)}\,.  
~~~~~~
q^{\dot \a}_{\dot -} = e^{-\half \phi + {i\over 2}
 H^{1} \mp {i\over 2} (H - \sqrt{3} x)}\,, 
$$

The relevant OPE's are 
$$
q^{1}_{+}(z) q^{2}_{+}(w) \sim {1\over (z-w)} e^{-\phi + i H^{1}}\,, ~~~~
q^{\dot 1}_{\dot -}(z) q^{\dot 2}_{\dot -}(w) 
\sim {1\over (z-w)} e^{-\phi + i H^{1}}\,, ~~~~
$$
$$
q^{1}_{-}(z) q^{2}_{-}(w) \sim {1\over (z-w)} e^{-\phi - i H^{1}}\,, ~~~~
q^{\dot 1}_{\dot +}(z) q^{\dot 2}_{\dot +}(w) 
\sim {1\over (z-w)} e^{-\phi - i H^{1}}\,, ~~~~
$$
\eqn\sugraD{
q^{1}_{+}(z) q^{\dot 1}_{\dot -}(w) \sim {1\over (z-w)^{1/2}} e^{-\phi + i H^{1} - i H}\,, ~~~~
q^{1}_{+}(z) q^{1}_{-}(w) 
\sim {1\over (z-w)^{3/2}} e^{-\phi}\,, 
}
$$
q^{1}_{+}(z) q^{\dot 1}_{\dot +}(w) \sim {\it reg.}\,, ~~~~~~~~~
q^{1}_{+}(z) q^{\dot 2}_{\dot -}(w) \sim (z-w)^{1/2} e^{- \phi + i H^{1} - i \sqrt{3} x} \sim 0 \,, 
$$
$$
q^{1}_{+}(z) q^{2}_{-}(w) \sim (z-w)^{1/2} e^{- \phi - i H - i \sqrt{3} x} \sim 0\,, ~~~~~
q^{1}_{+}(z) q^{\dot 2}_{\dot +}(w) \sim {1\over (z-w)} e^{- \phi -  i H}\,. 
$$
As before, the way to extract two set of charges which have no branching cuts 
is to consider $(\oint q^{\a}_{+}, \oint q^{\dot \a}_{\dot +})$ or 
$(\oint q^{\a}_{-}, \oint q^{\dot \a}_{\dot -})$. 
Such that the charges $\oint q^{\a}_{+}$ and $\oint q^{\dot \a}_{\dot +}$ 
represent the supersymmetry algebra in two dimensions and the OPE's between the two sets give a translation generators in the Liouville direction.  

The case D=5, n=2 is described by the 
following currents
$$
q^{\a}_{+} = e^{- \half \phi \mp {i\over 2}(H^{1} + H^{2}) 
- {i\over 2}(H + \sqrt{2} x)}\,, ~~~~
q^{\a'}_{+} = e^{- \half \phi \mp {i\over 2}(H^{1} - H^{2}) 
+ {i\over 2}(H + \sqrt{2} x)}\,, 
$$
$$
q^{\a}_{-} = e^{- \half \phi \pm {i\over 2}(H^{1} + H^{2}) 
+ {i\over 2}(H + \sqrt{2} x)}\,, ~~~~
q^{\a'}_{-} = e^{- \half \phi \pm {i\over 2}(H^{1} - H^{2}) 
- {i\over 2}(H + \sqrt{2} x)}\,, 
$$
\eqn\sugraE{
q^{\dot \a}_{\dot +} = e^{- \half \phi \pm {i\over 2}(H^{1} + H^{2}) 
- {i\over 2}(H - \sqrt{2} x)}\,, ~~~~
q^{\dot \a'}_{\dot +} = e^{- \half \phi \pm {i\over 2}(H^{1} - H^{2}) 
+ {i\over 2}(H - \sqrt{2} x)}\,, 
}
$$
q^{\dot \a}_{\dot -} = e^{- \half \phi \mp {i\over 2}(H^{1} + H^{2}) 
+ {i\over 2}(H - \sqrt{2} x)}\,, ~~~~
q^{\dot \a'}_{\dot -} = e^{- \half \phi \mp {i\over 2}(H^{1} - H^{2}) 
- {i\over 2}(H - \sqrt{2} x)}\,, 
$$
In the present case, out of these currents we select 
the four $(q^{a}_{+}, q^{a'}_{+})$. They are local w.r.t. each other and 
they close on the translation generators of the 4-dimensional 
transverse space.  Again, we can extend this set of 4 supercurrents by 
adding the currents $(q^{\dot a}_{\dot +}, q^{\dot a'}_{\dot +})$. 

\subsec{Construction of BRST charge D=1,n=0 case}

In order that the BRST charge is nilpotent and 
has the correct commutation properties with $G^{\prime -}$, 
we can rewrite $G^{\prime +}$ of \scA\ in a simpler form.
Working in the large Hilbert space, the BRST current can be 
expressed as follows 
\eqn\bbA{
G^{\prime +} = - e^{-R} e^{-\chi + 2(\phi + \kappa)} e^{R}\,, ~~~~~
R = \oint \Big[
(G^{+} + G^{-}) e^{\chi - \phi - \kappa} + {1\over 2} \p\phi 
e^{2 (\chi - \phi - \kappa)} \Big]\,.
}
where $G^{\pm}$ are the supersymmetry generators for the matter system 
given in \scC.\foot{Notice that the first term in $R$ 
can be written as follows $
G_{m} c \xi e^{-\phi}$. Therefore,  we can see that 
 $
(G_{m} c \xi e^{-\phi})(z) 
(G_{m} c \xi e^{-\phi})(w) \rightarrow (z-w)^{-1} 15 \p e^{-2\phi} \xi \p \xi c \p c$
which corresponds to the second term in $R$ up to an overall coefficient. 
} 

Using the above definitions \reA, it is easy to show that 
\eqn\bbB{
(G^{+} + G^{-}) = {1\over 2} \Big(
e^{i H} ( \p (\varphi - i x) - 2 i\p H) + e^{- iH} ( \p (\varphi + i x) + 2i \p H)
\Big)\,, 
}
and inserted into the combination of \bbA\ we get the 
three terms in $R$ expressed in terms of the 
new variables
\eqn\bbBA{
G^{+} e^{\chi - \phi - \kappa} = 
{1\over 2}\t^{+} \t^{\dot +} \Big(\p\varphi - i \p x' - 2 \p \omega - 2 \p \rho\Big) 
- i \p \t^{+} \t^{\dot +} 
}
$$
G^{-} e^{\chi - \phi - \kappa} = 
{1\over 2} 
e^{- 2 \omega} \Big(\p\varphi + i \p x' + 2 \p \omega + 2 \p \rho\Big)
+ i e^{-2 \omega} \t^{+} p_{+} 
\,.
$$
$$
\p \phi e^{2(\chi - \phi - \kappa)} = 
(-3 \p \t^{+} \t^{\dot +}  - \t^{+} \p\t^{\dot +}) e^{-2 \omega} + 
(2 \p \omega - 8 \p \rho - 4 i \p x') \t^{+} \t^{\dot +}  e^{-2 \omega}\,. 
$$
Finally, from \reA\ we can check that 
\eqn\bbD{
e^{-\chi + 2(\phi + \kappa)} = p_{\dot +}e^{\rho + \omega}\,.
}
from these definition we can deduce the form 
of the BRST operator by computing the similarity 
transformation on $p_{\dot +}e^{\rho + \omega}$.


\listrefs
\bye